\documentclass[%
 twocolumn,%
 aip,
 jcp,
 amsmath,amssymb,
 reprint,%
 groupedaddress
]{revtex4-1}

\usepackage[outercaption]{sidecap}
\usepackage{graphicx}
\usepackage{dcolumn}
\usepackage{bm}
\usepackage{color}
\usepackage{float}
\usepackage{epsfig}
\usepackage[a4paper,left=1.5cm,right=1.5cm, top=1.5cm,bottom=2cm]{geometry}

\newcommand{\be}{\begin{equation}}
\newcommand{\ee}{\end{equation}}
\newcommand{\bea}{\begin{eqnarray}}
\newcommand{\eea}{\end{eqnarray}}
\newcommand{\Fig}[1]{Fig.\,\ref{#1}}
\newcommand{\Eq}[1]{Eq.\,(\ref{#1})}
\newcommand{\etal}{{\it et al. }}
\newcommand{\la}{\langle}
\newcommand{\ra}{\rangle}
\newcommand{\nn}{\nonumber}

\newcommand{\RNum}[1]{\uppercase\expandafter{\romannumeral #1\relax}}

\begin{document}

\preprint{AIP/123-QED}

\title{Calculations of coherent two-dimensional electronic spectra using forward and backward stochastic wavefunctions}

\author{Yaling Ke}
\affiliation{
State Key Laboratory of Physical Chemistry of Solid Surfaces,
           Collaborative Innovation Center of Chemistry for Energy Materials,
           and Department of Chemistry, College of Chemistry and Chemical Engineering,
        Xiamen University, Xiamen, 361005, P. R. China
}%
\author{Yi Zhao}%
\email[E-mail: ]{yizhao@xmu.edu.cn}
\affiliation{
State Key Laboratory of Physical Chemistry of Solid Surfaces,
           Collaborative Innovation Center of Chemistry for Energy Materials,
           and Department of Chemistry, College of Chemistry and Chemical Engineering,
        Xiamen University, Xiamen, 361005, P. R. China
}%

\begin{abstract}
Within the well-established optical response function formalism, a new strategy with the central idea of employing the forward-backward stochastic Schr\"odinger equations in a segmented way to accurately obtain the two-dimensional (2D) electronic spectrum is presented in this paper. Based on the simple excitonically coupled dimer model system, the validity and efficiency of the proposed schemes are demonstrated in detail, along with the comparison against the deterministic hierarchy equations of motion and perturbative second-order time-convolutionless quantum master equations.  In addition, an important insight is provided in this paper that the characteristic frequency of the overdamped environment is an extremely crucial factor to regulate the lifetimes of the oscillating signals in 2D electronic spectra and of quantum coherence features of system dynamics. It is worth noting that the proposed scheme benefiting from its stochastic nature and wavefunction framework and many other advantages of substantially reducing the numerical cost, has a great potential to systematically investigate various quantum effects observed in realistic large-scale natural and artificial photosynthetic systems.

\end{abstract}


\maketitle

\section{Introduction}
Over the past two decades, heterodyne-detected two-dimensional electronic spectroscopy (2DES)\cite{Jonas-Annu.Rev.Phys.Chem.2003-p425,Brixner-J.Chem.Phys.2004-p4221,Brixner-Nature2005-p625,
Cho-Chem.Rev.2008-p1331,Abramavicius-Chem.Rev.2009-p2350}, generally being portrayed as the absorption frequency and emission frequency correlation plots at a sequence of populations times, has shown its great superiority in carrying an extremely abundant content about the structural properties and the dynamic information of various pigment-protein complexes and organic molecular aggregates\cite{Brixner-Nature2005-p625,Engel-Nature2007-p782,
Dostal-J.Am.Chem.Soc.2012-p11611,Westenhoff-J.Am.Chem.Soc.2012-p16484,Bennett-J.Am.Chem.Soc.2013-p9164,Collini-Science2009-p369,Abramavicius-J.Chem.Phys.2010-p11,
Panitchayangkoon-Proc.Natl.Acad.Sci.USA2010-p12766,Duan-Proc.Natl.Acad.Sci.USA2017-p8493,Maiuri-Chem2018-p20}. For instance, the observation of cross-peaks at short times and their oscillating behaviors provide the direct evidence of significant interchromophoric couplings. The environmental fluctuation features are imprinted onto peak broadening shapes and directions. The energy relaxation routes or the charge-transfer channels can also be dug up from these consecutive snapshots\cite{Brixner-Nature2005-p625,Engel-Nature2007-p782,
Dostal-J.Am.Chem.Soc.2012-p11611,Westenhoff-J.Am.Chem.Soc.2012-p16484,Lewis-J.Phys.Chem.Lett.2012-p503,Bennett-J.Am.Chem.Soc.2013-p9164,Collini-Science2009-p369,
Panitchayangkoon-Proc.Natl.Acad.Sci.USA2010-p12766,Duan-Proc.Natl.Acad.Sci.USA2017-p8493,Maiuri-Chem2018-p20}. The ultimate goal is to establish a quantitative relationship between the experimental 2DES signals and the ultrafast system dynamics. Recently, quantum state and process tomography has been proposed to extract the process matrix from two-color photon-echo signals for the dimer model system\cite{Yuen-Zhou-J.Chem.Phys.2011-p134505,Yuen-Zhou-Proc.Natl.Acad.Sci.USA2011-p17615}. Nevertheless, this step towards increasingly sophisticated molecular aggregates is still filled with a considerable amount of difficulties. In these systems, the electronic energy levels are often closely spaced and thus many peaks are obscured by further broadening, inhibiting the unambiguous assignment and ensuing tracking of all the desired signals. Besides, every point in the 2DES map stems from the dedicate interplays of a host of fundamental processes, intrinsically intensifying the complexity of interpretations. The most straightforward example of this fact is that the extensively-studied and relatively simple multichromophoric Fenna-Matthews-Olson (FMO) complex is hitherto trapped at the center of vehement controversy\cite{Brixner-Nature2005-p625,Cho-J.Phys.Chem.B2005-p10542,Engel-Nature2007-p782,Hayes-FaradayDiscuss.2011-p459,
Read-Biophys.J.2008-p847,Ishizaki-Proc.Natl.Acad.Sci.U.S.A.2009-p17255,
Ishizaki-J.Phys.Chem.B2011-p6227,Wilkins-J.Chem.TheoryComput.2015-p3411,Thyrhaug-J.Phys.Chem.Lett.2016-p1653,
Duan-Proc.Natl.Acad.Sci.USA2017-p8493}.
In the earlier stages when the unexpectedly long-lived beating of cross peaks were first revealed\cite{Engel-Nature2007-p782}, this wave-like signature was considered to be of pure electronic character and has triggered a large number of studies of how this coherence effect is harnessed to achieve marvelous energy transfer efficiency\cite{Plenio-NewJ.Phys.2008-p113019,Mohseni-J.Chem.Phys.2008-p174106,Panitchayangkoon-Proc.Natl.Acad.Sci.USA2010-p12766,
Kreisbeck-J.Phys.Chem.Lett.2012-p2828}. But to date, it nearly arrives at the consensus that the vibrational coherence effects also manifest themselves in the 2D electronic  spectra\cite{Turner-J.Phys.Chem.Lett.2011-p1904,Yuen-Zhou-J.Chem.Phys.2012-p234501,Kreisbeck-J.Phys.Chem.B2013-p9380,Fuller-Nat.Chem.2014-p706,
Tiwari-Proc.Natl.Acad.Sci.US2012-p1203}. The debate tends to become acute when a very recent experimental and theoretical revisit to the FMO complex suggests that the electronic dephasing takes place quickly at around tens of femtoseconds, while the residual periodic vibrational modulations are too weak to assist energy transfer\cite{Duan-Proc.Natl.Acad.Sci.USA2017-p8493}. However, there also exist other voices announcing that the lifetime of electronic coherence fitted from 2D spectra are substantially underestimated\cite{Kreisbeck-J.Phys.Chem.B2013-p9380} and some particular vibronic modes being in resonance with excitonic energy gaps do help long-range energy transfer\cite{Kolli-J.Chem.Phys.2012-p174109,Tiwari-Proc.Natl.Acad.Sci.US2012-p1203,Chin-Nat.Phys.2013-p113}. In order to resolve these conundrums, we need a very efficient theoretical method capable of accurately and thoroughly producing the 2DES results that are in quantitative agreement with the experimental ones, such that we can pinpoint the reasonable system and bath parameters and then generate a comprehensive picture of every details of system dynamics.

There are roughly two categories of methods to obtain the 2D electronic spectra. The first kind is termed as non-perturbative treatment as the field-matter interaction is explicitly incorporated into the Hamiltonian to dictate the system dynamics evolution\cite{Seidner-J.Chem.Phys.1995-p3998,
Kato-Chem.Phys.Lett.2001-p329,Gelin-Chem.Phys.2005-p135,Gelin-J.Chem.Phys.2005-p164112,Egorova-J.Chem.Phys.2007-p74314,
Mancal-J.Chem.Phys.2006-p234504,Cheng-J.Phys.Chem.A2007-p9499,Gelin-Acc.Chem.Res.2009-p1290,Seibt-J.Chem.Phys.2009-p134318,
Zhang-J.Phys.Chem.Lett.2016-p4488,Leng-Chem.Phys.Lett.2017-p79}, rendering it compatible with the majority of quantum dynamics methodologies in the arsenal of open quantum theories\cite{May-2008-p,Breuer-Rev.Mod.Phys.2016-p21002,Vega-Rev.Mod.Phys.2017-p15001,Tanimura-J.Phys.Soc.Jpn.2006-p82001}. In this case, what one can directly obtain is the total optical polarization rather than its specific spatial components responsible for the realistic experimental detections. Henceforth, an additional extraction procedure is required, for example, the phase-cycling scheme where the desired signal is obtained as the combination of a set of original signals acquired by changing the phase angles of pulse sequences\cite{Seidner-J.Chem.Phys.1995-p3998,Kato-Chem.Phys.Lett.2001-p329,Seibt-J.Chem.Phys.2009-p134318,
Zhang-J.Phys.Chem.Lett.2016-p4488}. To facilitate its applications in the aggregate molecules, Gelin and colleagues have proposed a phase-matching approach\cite{Gelin-Chem.Phys.2005-p135,Gelin-J.Chem.Phys.2005-p164112,Egorova-J.Chem.Phys.2007-p74314,Gelin-Acc.Chem.Res.2009-p1290}, suggesting that given the weak-field condition and rotating wave approximation, only  three independent time propagations of auxiliary density matrices are sufficient as for the determination of photo-echo spectroscopy. These methods have the advantage of naturally and trivially covering all the pulse-overlap effects, which are especially important when the pulse duration is finite and comparable to the ultrafast system dynamics timescale. Besides, all the possible pathways are included and intertwined together to produce the interference signal, which can be both a blessing and a curse, because it impairs our ability to resolve every elementary underlying physical process and thus provide less intuitive theoretical guidance than do as the perturbative treatments.

In the perturbative scenarios, Mukamel and coworkers have established and developed an unified theoretical foundation for various types of optical spectroscopic techniques\cite{mukamel1999principles,Zhang-J.Chem.Phys.1998-p7763,
Mukamel-Annu.Rev.Phys.Chem.2000-p691,Mukamel-Chem.Rev.2004-p2073,
Abramavicius-Europhys.Lett.2007-p17005,Mukamel-Acc.Chem.Res.2009-p1207,Abramavicius-Chem.Rev.2009-p2350,
Abramavicius-J.Chem.Phys.2010-p11}. Within this framework, the signal field is expressed in terms of  the so-called optical response functions in the ideal impulsive limit where the pulse peaks possess the delta-spiked profile. The response function is usually further spanned as the sum of active Liouville space pathways, which provide an intuitive basis for the interpretations and in-depth analyses of the observed spectra\cite{mukamel1999principles}. A magnitude of methods have been employed for the calculations of the third-order response functions and then to obtain the 2D electronic spectra\cite{mukamel1999principles,Zhang-J.Chem.Phys.1998-p7763,
Mukamel-Annu.Rev.Phys.Chem.2000-p691,Mukamel-Chem.Rev.2004-p2073,Abramavicius-Chem.Rev.2009-p2350,
Yang-Chem.Phys.2002-p163,Cho-J.Phys.Chem.B2005-p10542,
Chen-J.Chem.Phys.2010-p24505,Tanimura-J.Chem.Phys.2012-p22,
Hein-NewJ.Phys.2012-p23018,Ding-J.Chem.Phys.2011-p164107} . Nevertheless, the photosynthetic molecular aggregates intrinsically put a very high demand on both the accuracy and efficiency of the approaches\cite{Green-2003-p,Croce-2018-p}. On the one hand, the excitonic couplings, system-bath couplings, and the bath response timescales all reside in the similar levels, which is generally going beyond the valid parameter regimes of the second-order perturbative quantum master equations or their variants\cite{Yang-Chem.Phys.2002-p163,Ishizaki-J.Chem.Phys.2009-p234110,Chen-J.Chem.Phys.2010-p24505,Duan-Phys.Rev.E2015-p42708}. On the other hand, after the interrogations by three laser pulses, the system has a non-negligible probability to populate in higher-lying excited states whose number scales at least quadratically with the number of chromophores. Furthermore, the inclusion  of static disorders constitutes a key ingredient for the proper theoretical description of inhomogeneous peak broadening in 2D electronic spectra, which is realized by the repetitive calculations with different system parameters like the site energies, being sampled from a predefined distribution. These two aspects pose an extraordinarily heavy burden on the numerically exact methods, like hierarchy equations of motion\cite{Chen-J.Chem.Phys.2010-p24505,Tanimura-J.Chem.Phys.2012-p22,Hein-NewJ.Phys.2012-p23018,Kreisbeck-J.Phys.Chem.B2013-p9380,Ding-J.Chem.Phys.2011-p164107,
Dijkstra-J.Chem.Phys.2015-p212423} (HEOM) and quasi-adiabatic path-integral propogator approach\cite{Duan-Phys.Rev.E2015-p42708} (QUAPI) to conduct a systematic study even for the relatively simple FMO complex. Therefore, it is the motivation of this paper to present a novel numerical strategy for the accurate and efficient simulations of 2D electronic spectra for those photosynthetic complexes.

Very recently, several groups have been working on developing stochastic Schr\"odinger equations (SSEs) and optimizing their performance for computing the quantum dynamics and various linear response functions of large-scale complex systems\cite{Diosi-Phys.Rev.A1998-p1699,Roden-Phys.Rev.Lett.2009-p58301,Ritschel-NewJ.Phys.2011-p113034,Suess-Phys.Rev.Lett.2014-p150403,Song-J.Chem.Phys.2016-p224105,
Ke_J.Chem.Phys.2016_p24101,Ke-J.Chem.Phys.2017-p174105,Ke-J.Chem.Phys.2017-p184103,Link-Phys.Rev.Lett.2017-p180401}. In this paper, we would demonstrate the scheme of making use of stochastic wavefunctions to calculate the nonlinear optical spectra.
First of all, the response function $\mathcal{R}$ corresponding to a specific Liouville space pathway is formulated in the language of Feynman path-integral formalism by tracing over all the harmonic bath degrees of freedom. Afterwards, by introducing two correlated stochastic Gaussian processes ($\bm{\xi}^L(t),\bm{\xi}^R(t)$) with the detailed expression provided in the main context, we are able to unravel the response function as an ensemble average of the stochastic forward-backward stochastic wavefunctions, i.e., $\mathcal{R}=\left\la\psi^{L}{\psi^{R}}^*\right\ra_{\bm{\xi}^L\bm{\xi}^R}$. The time propagations of these stochastic wavefunctions, starting with the initial population in the common ground state, are achieved using the hierarchical or perturbative form of stochastic Schr\"odinger equations developed previously in our group\cite{Ke_J.Chem.Phys.2016_p24101,Ke-J.Chem.Phys.2017-p184103}, and will also be elaborately illustrated in the next section how they are exploited in the calculations of 2D electronic spectra. Owing to the Frenkel exciton model widely-used in the investigations of excitation energy transfer processes, the relaxation between different excitonic manifolds wouldn't occur during the time intervals in between two impulsive pulses. As a consequence, in our approach, the massive doubly excited states are only enrolled in the forward wavefunctions of excited state absorption pathways during the time intervals between the third and the fourth pulses, which greatly reduce the computational cost. Besides, the stochastic nature of the proposed method allows for the inclusion of static disorders in a quite trivial way. Another important point is that the stochastic wavefunctions are far more scaling-favorable than the density matrix in the practical simulations. In virtue of these advantages, the proposed schemes should hold great promise to accurately and efficiently simulate 2D electronic spectra of large-scale photosynthetic systems and then answer the on-going disputations.

The structure of the paper is arranged as follows: The theoretical model and  the technical details about using the forward-backward stochastic wavefunctions to simulate the 2D electronic spectra are presented in Sec. \ref{sec2}. In Sec. \ref{sec3}, the numerical results for the simple molecular dimer obtained from the hierarchical and perturbative forms of stochastic Schr\"odinger equations, as well as their deterministic analogs HEOM and time-convolutionless second-order quantum master equations, are compared to demonstrate the feasibility and validity of the proposed schemes. Incidently, it is also revealed in this section that the fastest electronic decoherence happens when the overdamped bath characteristic frequency is in resonance with the excitonic gaps. In addition, we have demonstrated how to implement the rotational average within the proposed methodology and use it to strengthen some specific Liouville pathways by varying the design of pulse sequence configurations. At last, the concluding remarks are given in Sec. \ref{sec4}.

\section{\label{sec2}Method}
It is well established that Frenkel exciton model is suitable for delineating the excitonic dynamics in the molecular aggregate composed of $N$ constitutent units, where each has two electronic levels, the ground state $|g_n\ra$ and the first excited state $|e_n\ra$ along with the vertical excitation energy $\epsilon_n$. In order to properly describe the two-dimensional optical spectra, three excitonic bands are typically involved, including the common ground state $|0\ra=\prod_{n=1}^N|g_n\ra$, single excitation manifold consisting of $N$ states $|n\ra=|e_n\ra\prod_{m\neq n}|g_m\ra$, and double excitation manifold with $N(N-1)/2$ doubly excited states $|nm\ra=|e_n\ra|e_m\ra\prod_{k\neq n,m}|g_k\ra \, (n<m)$. The excitonic Hamiltonian $\hat{H}_{ex}$ is then expressed as (For convenience, $\hbar =1$ and $k_B=1$ are adopted hereafter.)
\be
\label{ex_hamiltonian}
\begin{split}
\hat{H}_{ex}=&\sum_{n=1}^N \epsilon_n|n\ra\la n|+\sum_{n \neq m} V_{nm} |n\ra \la m| \\
& +\sum_{n<m,k<l} J_{nm,kl}|nm\ra \la kl|,
\end{split}
\ee
where $V_{nm}$ are the intermolecular couplings and the Hamiltonian matrix elements within the double-excitation subspace are given by
\be
\begin{split}
J_{nm,kl}=&\delta_{nk}\delta_{ml}(\epsilon_n+\epsilon_m) \\
&+\delta_{nk}(1-\delta_{ml})V_{ml}+\delta_{nl}(1-\delta_{mk})V_{mk} \\
&+\delta_{mk}(1-\delta_{nl})V_{nl}+\delta_{ml}(1-\delta_{nk})V_{nk}.
\end{split}
\ee
The states belonging to different manifolds  can only be accessed through the optical transitions described by the interaction term
\be
\hat{H}_{int}(t)=-\hat{\bm{\mu}}\cdot \bm{E}(\bm{r},t).
\ee
Here $\bm{E}(\bm{r},t)$ is the time-dependent semiclassical electric field, and $\hat{\bm{\mu}}$ denoting the total transition dipole operator reads
\be
\begin{split}
\hat{\bm{\mu}}=&\sum_{n=1}^N \bm{\mu}_{n}^{gs} \left(|0\ra \la n|+|n\ra \la 0|\right)\\
&+\sum_{n,k<l}\bm{\mu}_{n,kl}^{sd}\left(|n\ra\la kl|+|kl\ra \la n|\right),
\end{split}
\ee
with $\bm{\mu}_n^{gs}$ being the transition dipole moment of individual monomer and $\bm{\mu}_{n,kl}^{sd}=\delta_{nk}\bm{\mu}_{l}^{gs}+\delta_{nl}\bm{\mu}_{k}^{gs}$. Note that Franck-Condon approximation is invoked here, that is, $\bm{\mu}_n$ are independent of the vibrational degrees of freedom. Nevertheless, the peak shape broadening mechanisms are primarily attributed to these environmental degrees of freedom, which are represented by a multimode bath model consisting of numerous independent harmonic oscillators,
\be
\hat{H}_{ph}=\sum_{n=1}^N\sum_{\alpha}\frac{\hat{p}_{n\alpha}^2}{2}+\frac{\omega_{n\alpha}^2}{2}\hat{q}_{n\alpha}^2,
\ee
where $\hat{p}_{n\alpha}$ and $\hat{q}_{n\alpha}$ are mass-weighted momentum and coordinate operators of the bath mode with frequency $\omega_{n\alpha}$. In terms of macroscopic polarization calculations, this model is equivalent to another widely-used effective Brownian oscillator model\cite{mukamel1999principles,Mukamel-Chem.Rev.2004-p2073}, whereas the former is more commonly used in most of the open quantum dynamics approaches\cite{May-2008-p,Breuer-Rev.Mod.Phys.2016-p21002,Vega-Rev.Mod.Phys.2017-p15001}.
For simplicity, we have only taken into account the site excitation energy fluctuations  which are induced by their own bath. Henceforth, the coupling between the excitonic system and the bath can be explicitly written as
\be
\hat{H}_{ex-ph}=-\sum_{n=1}^N\sum_{\alpha}\omega_{n\alpha}^2d_{n\alpha}\hat{q}_{n\alpha}\hat{x}_n,
\ee
where $d_{n\alpha}$ is a mass-weighted position displacement, projected onto the $\hat{q}_{n\alpha}$ axis,  for the minimum of the potential energy surface of $|e_n\ra$ state  with respect to that of $|g_n\ra$  state, and it characterizes the system-bath coupling strength. In addition, the coupling operator is denoted by
\be
\hat{x}_n=|n\ra\la n|+\sum_{m<n}|mn\ra\la mn|+\sum_{m>n} |nm\ra\la nm|,
\ee
which implies that the energy fluctuations between the single and double excitation manifolds are cross-correlated\cite{Rancova-J.Phys.Chem.B2018-p1348}. As a matter of fact, the bath not simply leads to the decoherence of quantum features and the dissipation of excitation energies, but might also play a critical role in shaping the energy landscape resulting in an efficient energy transfer network\cite{Plenio-NewJ.Phys.2008-p113019,Mohseni-J.Chem.Phys.2008-p174106}.

We can write down  the total Hamiltonian that dictates the matter dynamics as
\be
\begin{split}
\hat{H}_{tot}&=\hat{H}_{ex}+\hat{H}_{ph}+\hat{H}_{ex-ph}+\hat{H}_{int}\\
&=\hat{H}_{mol}+\hat{H}_{int}.
\end{split}
\ee
One key quantity concerning to the statements of this paper is the macroscopic optical polarization defined as $\bm{P}(t)=\mathrm{tr}\{\hat{\bm{\mu}}\hat{\rho}(t)\}$, where $\hat{\rho}(t)$ is the time-evolving material density operator obtained by solving the following Liouville equation,
\be
i\frac{\partial }{\partial t}\hat{\rho}(t)=\left[\hat{H}_{tot},\hat{\rho}(t)\right].
\ee
By convention, one can always expand $\hat{\rho}(t)$ in power series of the electric field as per perturbation theory\cite{mukamel1999principles}, i.e., $\hat{\rho}(t)=\sum_{n=0}^{\infty} \hat{\rho}^{(n)}(t)$ when the incident electric field strengths are weak. In particular, the third-order nonlinear polarization $\bm{P}^{(3)}(t)=\mathrm{tr}\{\hat{\bm{\mu}}\rho^{(3)}(t)\}$ is closely related to various four-wave mixing experiments, where the first three non-collinear laser pulses interact with the material samples with the pulse central times  being $t_1$, $t_2$, and $t_3$, respectively. The explicit form of the electric field is generally given by
\be
\bm{E}(\bm{r},t)=\sum_{a=1}^3 \bm{e}_a \kappa_a E_a(t-t_a)e^{i\bm{k}_a\cdot \bm{r}-i\tilde{\omega}_a t}+c.c. ,
\ee
with $\bm{e}_a$, $\kappa_a$, $\bm{k}_a$, $\tilde{\omega}_a$ being the polarization unit vectors, field amplitudes, wave-vectors, and carrier frequencies of individual pulses. In what follows, we will focus on the case in which all these incident pulse durations are far shorter than their time intervals and any system dynamics of interest, such that the temporal envelops $E_a(t)$ can be reasonably assumed as $\delta$ functions, also termed as the impulsive limit.  In this situation, an ultrashort pulse has quite a broad span in frequency domain to cover all the optically allowed excitations, which in a sense provides instructive information for the finite duration cases.
The fourth pulse serves as a local oscillator to heterodyne detect a particular component of the complex-valued signal field $\bm{E}_{\bm{k}_s}$ along one of the following phase matching directions $\bm{k}_s=\pm\bm{k}_1 \pm \bm{k}_2 \pm \bm{k}_3$, which is determined according to the specific geometry of the experiment. Within the impulsive optical response function formalism, the measured signal field can be written as
\be
\bm{E}_{\bm{k}_s}(t,T,\tau)\propto i \bm{P}^{(3)}_{\bm{k}_s}(t)
=\kappa_1\kappa_2\kappa_3\mathcal{R}_{\bm{k}_s}(t,T,\tau)\bm{e}_1\bm{e}_2\bm{e}_3.
\ee
The time delays $\tau=t_2-t_1$, $T=t_3-t_2$, and $t=t_4-t_3$ ($t_4 $ is the detection moment) are frequently referred to as coherence time, population (or waiting) time, and rephasing time, respectively. The term $\mathcal{R}_{\bm{k}_s}(t,T,\tau)$ in the above equation is the so-called material third-order response function to the external field, which constitutes an important basis for systematic and in-depth analyses of a vast number of nonlinear optical spectroscopies. $\mathcal{R}_{\bm{k}_s}(t,T,\tau)$ is actually a fourth-rank tensor, but for the sake of clear demonstration, we have temporally neglected its tensorial nature, and will revisit this issue in next section when treating orientational averaging of an isotropic sample. Under the rotating wave approximation (RWA), only three components, called rephasing, non-rephasing, and double quantum coherence terms, survive and radiate into the directions  $\bm{k}_{\RNum{1}}=-\bm{k}_1+\bm{k}_2+\bm{k}_3$, $\bm{k}_{\RNum{2}}=\bm{k}_1-\bm{k}_2+\bm{k}_3$, and $\bm{k}_{\RNum{3}}=\bm{k}_1+\bm{k}_2-\bm{k}_3$, respectively. Basically, every component can be further dissected into a few Liouville space pathways. For instance, there are both three Liouville pathways contributing to the rephasing and nonrephasing response functions,
\be
\label{rephasing_response}
\begin{split}
\mathcal{R}_{\bm{k}_{\RNum{1}}}(t,T,\tau)=&\mathcal{R}_{\bm{k}_{\RNum{1}}}^{GSB}(t,T,\tau)+\mathcal{R}_{\bm{k}_{\RNum{1}}}^{SE}(t,T,\tau)\\
&-\mathcal{R}_{\bm{k}_{\RNum{1}}}^{ESA}(t,T,\tau),
\end{split}
\ee
and
\be
\label{nonrephasing_response}
\begin{split}
\mathcal{R}_{\bm{k}_{\RNum{2}}}(t,T,\tau)=&\mathcal{R}_{\bm{k}_{\RNum{2}}}^{GSB}(t,T,\tau)+\mathcal{R}_{\bm{k}_{\RNum{2}}}^{SE}(t,T,\tau)\\
&-\mathcal{R}_{\bm{k}_{\RNum{2}}}^{ESA}(t,T,\tau).
\end{split}
\ee
The superscript $GSB$, $SE$, and $ESA$ are acronyms for ground state bleaching, stimulated emission, and excited state absorption, respectively, which provide intuitive hints about the underlying physical processes. Note that the double quantum coherence signal which is though helpful in revealing the double-excitation band structure and the correlation between different excitonic manifolds since only the optical coherence states take part into the two Liouville pathways therein\cite{Xu-Chin.J.Chem.Phys.2011-p497,Nemeth-J.Chem.Phys.2010-p94505}, is less competent in directly reflecting the energy relaxation features and exciton transport dynamics when compared to the other two components. Therefore, in the following context, we will concentrate on discussing about the 2D electronic photon-echo spectra which are obtained as the superposition of rephasing and nonrephasing signals, because these two parts in fact can be accessed within the same experimental setup by switching the time order of the first two pulses. More specifically, one generally obtain the 2D electronic spectra by fixing the population time $T$ at a certain value and then implementing the double Fourier-Laplace transformations upon $\mathcal{R}_{\bm{k}_{\RNum{1}}}$ and $\mathcal{R}_{\bm{k}_{\RNum{2}}}$ with respect to the coherence time $\tau$ and the rephasing time $t$ as\cite{Chen-J.Chem.Phys.2010-p24505}
\bea
\label{2D_spectra}
\nn S(\omega_{t},T,\omega_{\tau})=&\mathrm{Re}\int_0^{\infty}\mathrm{d}t\int_0^{\infty}\mathrm{d}\tau \left[e^{i(\omega_t t+\omega_{\tau} \tau)}\mathcal{R}_{\bm{k}_{\RNum{1}}}(t,T,\tau)\right.\\
&\left.+e^{i(\omega_t t-\omega_{\tau} \tau)}\mathcal{R}_{\bm{k}_{\RNum{2}}}(t,T,\tau)\right].
\eea

After decades of developments in this area, a great number of methodologies, ranging from analytical expressions to numerical approaches, have been put forward and then employed to calculate the time-domain response functions\cite{mukamel1999principles,Zhang-J.Chem.Phys.1998-p7763,
Mukamel-Annu.Rev.Phys.Chem.2000-p691,Mukamel-Chem.Rev.2004-p2073,
Abramavicius-Europhys.Lett.2007-p17005,Mukamel-Acc.Chem.Res.2009-p1207,Abramavicius-Chem.Rev.2009-p2350,
Abramavicius-J.Chem.Phys.2010-p11,Provazza-J.Chem.TheoryComput.2018-p856}, being in tight connection with the open quantum dynamics theories. Recently, the description of large-scale open quantum systems using non-Markovian stochastic Schr\"odinger equations has  attracted increasing attentions. SSEs have an advantageous scaling relationship with the system size over the density-matrix-based framework. Besides, it should be emphasized that, regarding the high-order optical response function simulations, the inclusion of static disorders has long been a bottleneck of many deterministic methods. Instead, one can readily consider the static and dynamical disorders simultaneously in propagating SSEs. In what follows, we will demonstrate how to use SSEs to calculate the nonlinear optical response functions.

\begin{figure*}
  \centering
  \includegraphics[width=0.8\textwidth]{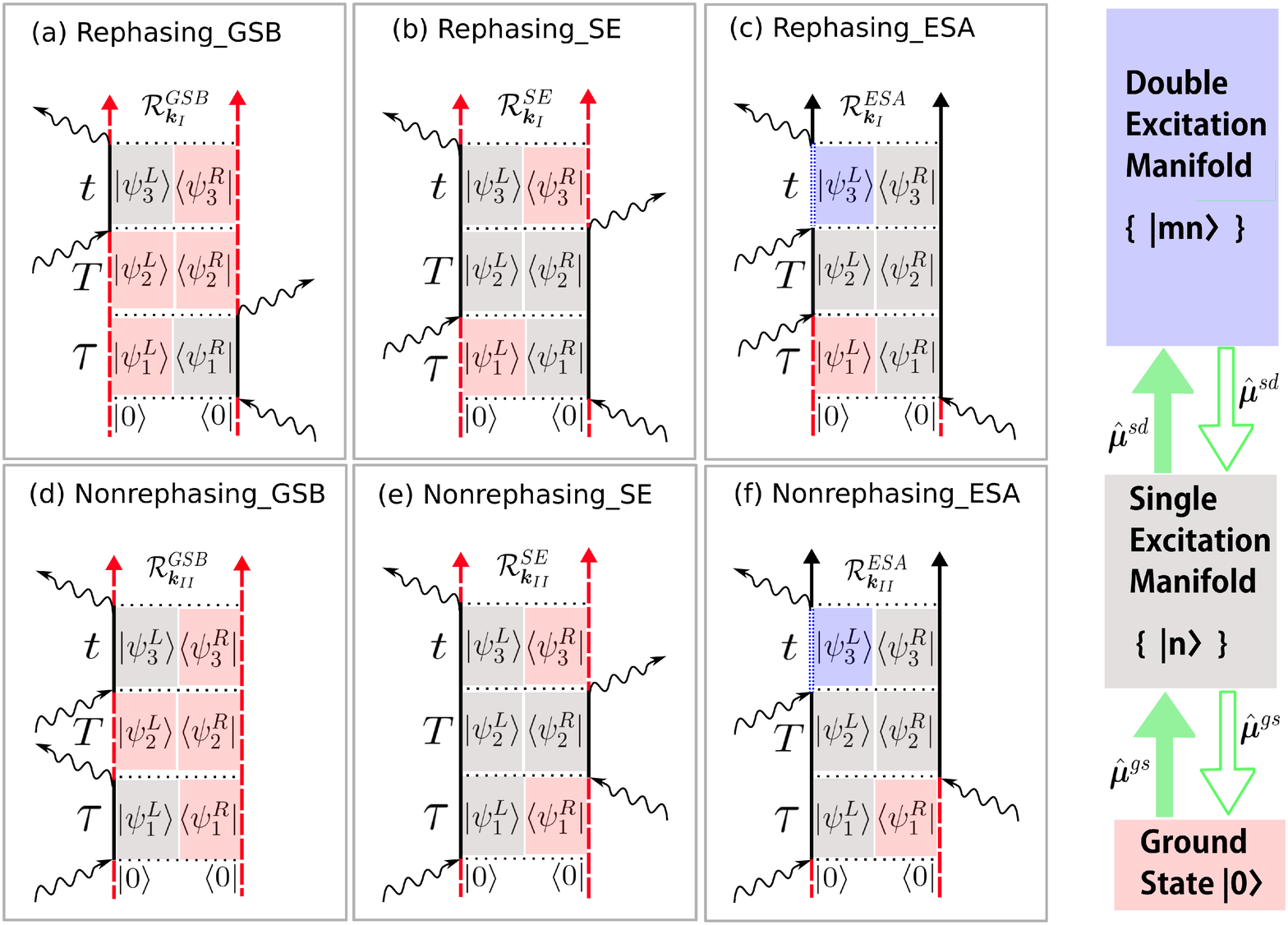}\\
  \caption{Schematic of double-sided Feynman diagrams for six different Liouville space pathways, as individually specified in the top of panels. $\tau$, $T$, and $t$ are the coherence, population, and rephasing time delays, respectively. The vertical straight lines with arrows from the bottom up represent the time evolution direction, denoting that $|\psi_1^{L/R}\ra$, $|\psi_2^{L/R}\ra$, and $|\psi_3^{L/R}\ra$ are propagated in a sequential way.  The ket and bra sides represent the forward and the backward stochastic wavefunctions, respectively. The boxes colored in red, grey, and blue mean that the wavefunctions therein are populated in common ground state, single excitation subspace, and double excitation subspace, respectively.  States belonging to different excitation manifolds are only switchable through the interaction with the electric fields, which is indicative of curves with arrows. }\label{Fig1}
\end{figure*}
Notice that the items in the r.h.s. of \Eq{rephasing_response} and \Eq{nonrephasing_response}, schematically shown as the double-sided Feynman diagrams in \Fig{Fig1}, can be uniformly formulated as
\begin{widetext}
\be
\label{general_response_function}
\mathcal{R}(t,T,\tau)=\mathrm{tr}\left\{ \hat{M}_4^L \left[ \mathcal{G}(t) \left( \hat{M}_3^L  \left[\mathcal{G}(T)\left( \hat{M}_2^L  \left[ \mathcal{G}(\tau)  \left(\hat{M}^L_1\hat{\rho}_0 \hat{M}^R_1\right) \right]\hat{M}_2^R \right) \right] \hat{M}_3^R  \right) \right] \right\}=\mathrm{tr}_{ex}\left\{\hat{R}_{ex}(t,T,\tau) \right\},
\ee
where $\hat{\rho}_0=|0\ra\la 0|e^{-\beta \hat{H}_{ph}}/\mathrm{tr_{ph}}\{e^{-\beta\hat{H}_{ph}} \}$ with $\beta$ being the inverse temperature, $\mathcal{G}(t)\hat{A}=e^{-i\hat{H}_{mol}t}\hat{A}e^{i\hat{H}_{mol}t}$, and $\hat{M}_j^{L/R}\,(j=1, 2, 3, 4)$ can be either unit operator $\hat{\bm{I}}$, $\hat{\bm{\mu}}^{gs}$, or $\hat{\bm{\mu}}^{sd}$, depending specifically on which Liouville pathway it represents. To proceed, we rewrite $\hat{R}_{ex}(t,T,\tau)$ defined in \Eq{general_response_function} using the path integral formalism\cite{Feynman-Ann.Phys.1963-p118},
\be
\label{response_function}
\begin{split}
\hat{R}_{ex}(t,T,\tau)=&\hat{M}_4^L \mathrm{tr_{ph}}\left\{ \mathcal{G}(t) \left( \hat{M}_3^L  \left[\mathcal{G}(T)\left( \hat{M}_2^L  \left[ \mathcal{G}(\tau)  \left(\hat{M}^L_1\hat{\rho}_0 \hat{M}^R_1\right) \right]\hat{M}_2^R \right) \right] \hat{M}_3^R  \right) \right\} \\
=&\hat{M}_4^L \int \mathcal{D}\bm{x}^+(s)\int \mathcal{D}\bm{x}^-(s) M^L_{3}(\bm{x}^+(T+\tau))  M^L_{2}\left[\bm{x}^+(\tau)\right] e^{iS_0(\bm{x}^+(s))} \\
&  \left(\hat{M}^L_1 |0\ra\la 0|\hat{M}^R_1\right)  e^{-iS_0(\bm{x}^-(s))}  M^R_{2}(\bm{x}^-(\tau))  M^R_{3}(\bm{x}^-(T+\tau))   \mathcal{F}(\bm{x}^+(s),\bm{x}^-(s)).
\end{split}
\ee
\end{widetext}
In the above expression, $\bm{x}^{+}(s)$ and $\bm{x}^{-}(s)$ are the forward and backward system paths, respectively. $S_0$ is the action functional originating from the excitonic Hamiltonian $H_{ex}$, and the Feynman-Vernon influence functional
\be\label{influence_functional}
\begin{split}
\mathcal{F}(&\bm{x}^+(s),\bm{x}^-(s))=\exp\Bigg\{-\sum_{n=1}^N\int_0^{t+T+\tau}\mathrm{d}s\int_0^{s}\mathrm{d}s' \big[x_n^+(s)\\
&-x_n^-(s)\big]\left[\alpha_n(s-s')x_n^+(s')-\alpha_n^*(s-s')x_n^-(s')\right]\Bigg\}
\end{split}
\ee
records the environmental impacts on the system dynamics of interest, after tracing over all the bath degrees of freedom.
The environmental dynamical information is statistically retained in the bath correlation function that reads
\be
\begin{split}
\alpha_n(t)&=\alpha_n^{\mathrm{re}}(t)+i\alpha_n^{\mathrm{im}}(t)\\
&=\int_0^{\infty}\frac{J_n(\omega)}{\pi}\left[ \coth\left(\frac{\beta \omega}{2}\right)\cos \omega t-i\sin \omega t \right]\mathrm{d}\omega,
\end{split}
\ee
and it intrinsically relies on the more fundamental temperature-free spectral density function
\be
J_n(\omega)=\frac{\pi}{2}\sum_{\alpha} \omega_{n\alpha}^3 d_{n\alpha}^2\delta(\omega-\omega_{n\alpha}),
\ee
which is a system-bath coupling weighted bath density distribution function. In the condense phase where the bath frequencies are dense, $J(\omega)$ turns to be continuous.

An important step leading to the eventual wavefunction-based formalism is to unravel the explicit entanglement between the forward path $\bm{x}^+(s)$ and the backward path $\bm{x}^-(s)$ in $\mathcal{F}(\bm{x}^+(s),\bm{x}^-(s))$. To this end, one impossible way is to adopt the Hubbard-Stratnovich transformation by introducing two correlated sequences of Gaussian stochastic processes\cite{Hubbard-Phys.Rev.Lett.1959-p77,Stratonovich-1957-p416,Stockburger-J.Chem.Phys.1999-p4983,Shao-J.Chem.Phys.2004-p5053,Ke_J.Chem.Phys.2016_p24101,Ke-J.Chem.Phys.2017-p184103},
$\bm{\xi}^L(s)=(\xi_1^L(s),\cdots,\xi_N^L(s))$ and  $\bm{\xi}^R(s)=(\xi_1^R(s),\cdots,\xi_N^R(s))$, with the following definition which is also its numerical generation schemes\cite{Ke-J.Chem.Phys.2017-p184103},
\begin{subequations}
\label{stochastic}
\be
\begin{split}
\xi^c_n(s)=\sum_{\alpha} & \sqrt{\frac{\omega_{n\alpha}^3d_{n\alpha}^2}{2\sinh\left(\frac{\beta\omega_{n\alpha}}{2}\right)}}\Big[\theta_{n\alpha}^1\cos\omega_{n\alpha}(s-i\beta/4)\\
&-\theta_{n\alpha}^2 \sin{\omega_{n\alpha}(s-i\beta/4)}\Big];
\end{split}
\ee
\be
\begin{split}
\xi^L_n(s)=\sum_{\alpha}& \frac{\sqrt{\omega_{n\alpha}^3d_{n\alpha}^2\tanh\left(\frac{\beta\omega_{n\alpha}}{4}\right)}}{2}
\Big[\theta_{n\alpha}^3\cos (\omega_{n\alpha}s) \\
&+\theta_{\alpha}^4\sin (\omega_{n\alpha}s)\Big]+\xi_n^c(s);
\end{split}
\ee
\be
\begin{split}
\xi^R_n(s)=\sum_{\alpha}& \frac{\sqrt{\omega_{n\alpha}^3d_{n\alpha}^2\tanh\left(\frac{\beta\omega_{n\alpha}}{4}\right)}}{2}
\Big[\theta_{n\alpha}^5\cos (\omega_{n\alpha}s) \\
&+\theta_{n\alpha}^6\sin (\omega_{n\alpha}s)\Big]+\xi_n^c(s),
\end{split}
\ee
\end{subequations}
where $\theta_{n\alpha}^i\,(i=1,\cdots,6)$ are independent Gaussian random variables with the average values $\la \theta_{n\alpha}^i\ra=0$ and the correlations $\la \theta_{n\alpha}^i\theta_{m\beta}^j \ra=\delta_{ij}\delta_{nm}\delta_{\alpha\beta}$.  It can be readily verified that $\xi_n^L(s)$ and $\xi_n^R(s)$ satisfy
\be
\la \xi_n^L(s) \xi_m^L(s')\ra=\la \xi_n^R(s) \xi_m^R(s')\ra=\delta_{nm}\alpha_n^{\mathrm{re}}(s-s')
\ee
and
\be
\la \xi_n^L(s) {\xi_m^R}^*(s')\ra=\delta_{nm}\alpha_n^*(s-s').
\ee
\begin{widetext}
As such, the following expression holds,
\be
\label{if_unravel}
\begin{split}
\mathcal{F}(\bm{x}^+(s),\bm{x}^-(s))=&\int \mathcal{D}^2\bm{\xi}^L \int \mathcal{D}^2{\bm{\xi}^R}^*P(\bm{\xi}^L,{\bm{\xi}^R}^*)e^{-i\sum_{n=1}^N\int_0^{t+T+\tau} \left[x_n^+(s)\xi_n^L(s)-x_n^-(s){\xi_n^R}^*(s)\mathrm{d}s\right]}\\
&e^{-i\sum_{n=1}^N\int_0^{t+T+\tau} \mathrm{d}s \int_0^{s}\mathrm{d}s'x_n^+(s)\alpha^{\mathrm{im}}_n(s-s')x_n^+(s')} e^{+i\sum_{n=1}^N\int_0^{t+T+\tau} \mathrm{d}s \int_0^{s}\mathrm{d}s'x_n^-(s)\alpha^{\mathrm{im}}_n(s-s')x_n^-(s')}.
\end{split}
\ee
By bringing \Eq{if_unravel} back into \Eq{response_function}, it yields,
\be
\label{sto_response}
\hat{R}_{ex}(t,T,\tau)=\int \mathcal{D}^2\bm{\xi}^L\int \mathcal{D}^2{\bm{\xi}^R}^* P(\bm{\xi}^L,{\bm{\xi}^R}^*) \hat{M}_4^L|\Psi^{L}(t,T,\tau;\bm{\xi}^L)\ra   \la \Psi^{R}(t,T,\tau;{\bm{\xi}^R}^*)|,
\ee
with
\be
\label{path_inte}
\begin{split}
|\Psi^{L/R}(t,T,\tau;\bm{\xi}^{L/R})\ra=\int \mathcal{D}\bm{x}(s)e^{iS_0(\bm{x}(s))}\Big\{& e^{-i\sum_{n=1}^N\int_0^{t+T+\tau} \mathrm{d}sx_n(s)\left[\xi_n^{L/R}(s)+ \int_0^{s}\mathrm{d}s'\alpha_n^{\mathrm{im}}(s-s')x_n(s')\right]}\\
&M^{L/R}_{3}(\bm{x}(T+\tau))M^{L/R}_{2}(\bm{x}(\tau))\Big\}\hat{M}^{L/R}_1 |0\ra.
\end{split}
\ee
The exponential term within the curly brackets can be expanded and rearranged into the form of an infinite and intricate sum, while for each component we make use of the identity
\be
\int_{\bm{x}_i}^{\bm{x}_f}\mathcal{D}\bm{x}(s)e^{iS_0(\bm{x}(s))}\bm{x}(t_n)\cdots \bm{x}(t_1)=\la \bm{X}_f|\mathcal{T}_{X}^{\leftarrow}\left\{\hat{\bm{X}}(t_n)\cdots \hat{\bm{X}}(t_1)\right\}|\bm{X}_i\ra,\, t_f>t_n,\cdots,t_1>t_i.
\ee
Here $\hat{\bm{X}}(t)=e^{i\hat{H}_{ex}t}\hat{\bm{x}}e^{-i\hat{H}_{ex}t}$, and $\mathcal{T}_X^{\leftarrow}$ represents the chronological time-ordering operation of the products $\hat{\bm{X}}(t_n)\cdots\hat{\bm{X}}(t_1)$. By commuting the time-ordering operation with the summation operation and then refolding them back into the exponential form, we can rewrite \Eq{path_inte} as
\be
\begin{split}
|\Psi^{L/R}(t,T,\tau;\bm{\xi}^{L/R})\ra=e^{-i\hat{H}_{ex}(t+T+\tau)}\mathcal{T}_{X}^{\leftarrow} \Big\{ & e^{-i\sum_{n=1}^N\int_0^{t+T+\tau}\mathrm{d}s \hat{X}_n(s)\left[\xi_n^{L/R}(s)+\int_0^{s}\mathrm{d}s'\alpha_n^{\mathrm{im}}(s-s')\hat{X}_n(s')\right]} \\
& M^{L/R}_3(\hat{\bm{X}}(T+\tau))  M^{L/R}_2(\hat{\bm{X}}(\tau))\Big\}\hat{M}^{L/R}_1 |0\ra.
\end{split}
\ee
\end{widetext}
For simplicity, the superscript $L/R$ will be omitted in the following derivations. Benefiting from the impulsive limit approximation, it would be numerically preferable to introduce a set of piecewise functions\cite{Ishizaki-Chem.Phys.2008-p185}:
\be
\label{wavefunctions}
|\psi^{\bm{h}}(s,\bm{\xi})\ra=\left\{
\begin{array}{lc}
|\psi_1^{\bm{h}}(s,\bm{\xi})\ra,& 0<s<\tau \\
|\psi_2^{\bm{h}}(s,\bm{\xi})\ra,& \tau<s<\tau+T \\
|\psi_3^{\bm{h}}(s,\bm{\xi})\ra,& \tau+T<s<\tau+T+t \\
\end{array}
\right.
\ee
with the index vector $\bm{h}=(h_1,h_2,\cdots,h_N)$ where $h_n$ are non-negative integers, and the explicit expressions for $|\psi_1^{\bm{h}}\ra$, $|\psi_2^{\bm{h}}\ra$, and $|\psi_3^{\bm{h}}\ra$ are consistently given by
\label{auxiliary_wavefunctions}
\be
\begin{split}
 |\psi_{j}^{\bm{h}}(s;\bm{\xi})\ra=&e^{-i\hat{H}_{ex}s}\mathcal{T}_{X}^{\leftarrow} \bigg\{ e^{-i\sum_{n=1}^N\int_0^{s}\mathrm{d}s'  \hat{X}_n(s')\xi_n(s')\mathrm{d}s'}\\
 &e^{-i\sum_{n=1}^N\int_0^{s}\mathrm{d}s'\int_0^{s'}\mathrm{d}s''\hat{X}_n(s')\alpha_n^{\mathrm{im}}(s'-s'')\hat{X}_n(s'')}\\
& \prod_{n=1}^N \left(\int_0^{s} \alpha_{n}^{\mathrm{im}}(s-s')\hat{X}_n(s')\mathrm{d}s' \right)^{h_n}\\
 &M_3^{b_j}(\hat{\bm{X}}(\tau+T))M_2^{a_j}(\hat{\bm{X}}(\tau))\bigg\} \hat{M}_1 |0\ra.
\end{split}
\ee
We would acquire $ |\psi_{1}^{\bm{h}}\ra$ when $(a_1,b_1)=(0,0)$, $ |\psi_{2}^{\bm{h}}\ra$ when $(a_2,b_2)=(1,0)$, and $ |\psi_{3}^{\bm{h}}\ra$ when $(a_3,b_3)=(1,1)$. Besides, $|\psi_{3}^{\bm{h}}\ra$ equals to $|\Psi(t,T,\tau,\bm{\xi})\ra$ when $\bm{h}=\bm{0}$. In a special case where every bath can be represented by an overdamped Brownian oscillator, its spectral density function is of the form
\be
\label{Debye}
J_n(\omega)=\frac{2\lambda_n\omega\gamma_n}{\omega^2+\gamma_n^2},
\ee
such that the imaginary part of bath correlation function appears as a simple exponential decay function
\be
\alpha_{n}^{\mathrm{im}}(t)=-\lambda \gamma e^{-\gamma t}.
\ee
Then, the time-derivation of \Eq{wavefunctions} has proven to produce a hierarchical structure of self-contained differential equations\cite{Ke_J.Chem.Phys.2016_p24101}
\be
\label{hierarchy}
\begin{split}
i\frac{\partial }{\partial s}&|\psi^{\bm{h}}(s;\bm{\xi}) \ra =\Bigg[\hat{H}_{ex}+\sum_{n=1}^N\left(\xi_n(s)\hat{x}_n -ih_n\gamma_n\right) \Bigg] |\psi^{\bm{h}}(s;\bm{\xi}) \ra  \\
& +\sum_{n=1}^N\left(\hat{x}_n |\psi^{\bm{h}_n^+}(s;\bm{\xi}) \ra-ih_n\lambda_n\gamma_n\hat{x}_n|\psi^{\bm{h}_n^-}(s;\bm{\xi})\ra\right),
\end{split}
\ee
where $\bm{h}_n^{\pm}=(h_1,\cdots, h_n\pm 1, h_N)$. The initial condition is prepared as $|\psi^{\bm{0}}(0;\bm{\xi}) \ra=|\psi_1^{\bm{0}}(0;\bm{\xi}) \ra=\hat{M}_1|0\ra$, and $|\psi^{\bm{h}}(0;\bm{\xi}) \ra=|\psi_1^{\bm{h}}(0;\bm{\xi}) \ra=0\, (\bm{h}\neq \bm{0})$.

\Eq{hierarchy} is henceforth referred to as the hierarchical form of stochastic Schr\"odinger equations (HSSE).
In the numerical implementations, the hierarchy is indispensably truncated as a certain depth, at which the results are ensured to be converged.
$|\psi_1^{\bm{h}}(s,\bm{\xi})\ra$ are obtained by integrating the above truncated differential equation till the moment $\tau$, and then it is impinged by the operator $\hat{M}_2$ to yield $|\psi_2^{\bm{h}}(\tau,\bm{\xi})\ra=\hat{M}_2|\psi_1^{\bm{h}}(\tau,\bm{\xi})\ra$ for all $\bm{h}$. As the time evolves, $|\psi_2^{\bm{h}}(s,\bm{\xi})\ra$ are propagated to another time point $s=\tau+T$, at which we perform $|\psi_3^{\bm{h}}(\tau+T,\bm{\xi})\ra=\hat{M}_3|\psi_2^{\bm{h}}(\tau+T,\bm{\xi})\ra$.
Likewise, $|\psi_3^{\bm{h}}(\tau+T+t,\bm{\xi})\ra$ are obtained for a time delay $t$ propagation using \Eq{hierarchy}.  This successive  procedure to obtain the optical response functions for six different Liouville pathways is schematically shown in \Fig{Fig1}.

Noting that there is an implicit benefit of numerical scheme \Eq{wavefunctions}, $|\psi_1^{\bm{h}}\ra$, $|\psi_2^{\bm{h}}\ra$, and $|\psi_3^{\bm{h}}\ra$ can be the vectors blocked inside a specific kind of excitation manifold instead of the whole excitonic space. Taking the ground state bleaching pathway contributing to the non-rephasing signals as an example, as clearly shown in \Fig{Fig1} (d), $|\psi_1^{\bm{h},L}(s)\ra$  and $|\psi_3^{\bm{h},L}(s)\ra$ belong to the single excitation subspace, while the rest $|\psi_2^{\bm{h},L}(s)\ra$, $|\psi_1^{\bm{h},R}(s)\ra$, $|\psi_2^{\bm{h},R}(s)\ra$, and $|\psi_3^{\bm{h},R}(s)\ra$ stay in the ground state and are thus kept constant within the corresponding time intervals. It is worth mentioning that the considerable amount of doubly excited states only participate in $|\psi_3^{\bm{h},L}(s)\ra$ of the excited state absorption pathways, as illustrated in \Fig{Fig1} (c) and (f) for rephasing and nonrephasing signals, respectively. In this regard, this practice can no doubt substantially save numerical resources when dealing with large size systems.

The 2D electronic spectra obtained through \Eq{hierarchy} are numerically exact. Moreover, the extension of HSSE {\color{red}to the cases} where the bath is constructed by a few distinct vibrational modes or other continuous spectral density functions  is available with the aid of some decomposition schemes\cite{Meier-J.Chem.Phys.1999-p3365,Liu_J.Chem.Phys._2014_p134106,
Ritschel-J.Chem.Phys.2015-p34115,Ke_J.Chem.Phys.2016_p24101}. But in these cases, the equations generally take on a more convoluted look and the numerical requirements become more  demanding as well. Given the fact that in comparison with the direct system dynamics simulations, the calculation of the third-order response functions actually remains an extremely time-consuming task even for small-size systems. As such, when the interplay between the system and bath is not very strong which is typical for most photosynthetic complexes, it is favorable and rational to further invoke some approximations. We have sought for a perturbative solution on the basis of the above-mentioned hierarchical scheme. Note that the application of hierarchy-based approaches to large-scale molecular aggregates is generally limited by a great number of auxiliary wavefunctions. Nevertheless, in the end, only the zeroth order wavefunction $|\psi^{\bm{0}}\ra$ is what we want. Such that, the key of the adaption is to close \Eq{hierarchy} at its lowest order with a particular truncation scheme shown below, which in essence is a second-order perturbation approximation about system-bath coupling\cite{Ke-J.Chem.Phys.2017-p184103},
\begin{widetext}
\be
\label{approximation}
\begin{split}
 |\psi_j^{\bm{1}_m}(s,\bm{\xi})\ra = &e^{-i\hat{H}_{ex}s}\mathcal{T}_{X}^{\leftarrow} \bigg\{ e^{-i\sum_{n=1}^N\left[\int_0^{s} \hat{X}_n(s')\xi_n(s') \mathrm{d}s'+\int_0^{s} \mathrm{d}s' \int_0^{s'}\mathrm{d}s''\hat{X}_n(s')\alpha_n^{\mathrm{im}}(s'-s'')\hat{X}_n(s'')\right]} \\
 & M_3^{b_j}(\hat{\bm{X}}(T+\tau))M_2^{a_j}(\hat{\bm{X}}(\tau)) \left(\int_0^{s} \alpha_m^{\mathrm{im}}(s-s')\hat{X}_m(s')\mathrm{d}s' \right) \bigg\}\hat{M}_1|0\ra\\
 \approx&\left(\int_0^{s} \alpha_{m}^{\mathrm{im}}(s')e^{-i\hat{H}_{ex}s'}\hat{x}_me^{i\hat{H}_{ex}s'}\mathrm{d}s' \right)
 |\psi_j^{\bm{0}}(s,\bm{\xi})\ra,
\end{split}
\ee
\end{widetext}
where $\bm{1}_m$ means $h_m=1$, $h_{k}=0$ for $k\neq m$. Subsequently, we substitute  all $|\psi_j^{\bm{1}_m}(s,\bm{\xi})\ra$ in \Eq{hierarchy} ($m$ runs from 1 to N) with the approximate expression in \Eq{approximation}, such that the differential equation for  $|\psi^{\bm{0}}(s,\bm{\xi})\ra$ turns into
\be
\label{perturbation}
\begin{split}
i\frac{\partial }{\partial s}|\psi^{(\bm{0})}(s;\bm{\xi}) \ra =&\left[\hat{H}_{ex}+\sum_{n=1}^N\xi_n(s)\hat{x}_n\right] |\psi^{(\bm{0})}(s;\bm{\xi}) \ra\\
&+  \sum_{n=1}^N\int_0^{s}\alpha_{n}^{\mathrm{im}}(s')\Big[\hat{x}_n e^{-i\hat{H}_{ex}s'} \\
&\qquad \hat{x}_ne^{i\hat{H}_{ex}s'}\mathrm{d}s'\Big]|\psi^{(\bm{0})}(s;\bm{\xi})\ra,
\end{split}
\ee
which is called as time-local perturbative stochastic Schr\"odinger equations (PSSE). It should be highlighted that, PSSE is straightforwardly applicable to arbitrary kind of spectral density function, which is increasingly important since the realistic environments are more likely to be highly-structured\cite{Lee-J.Phys.Chem.Lett.2016-p3171,Padula-J.Phys.Chem.B2017-p10026}. In addition, it is found in our previous studies\cite{Ke-J.Chem.Phys.2017-p184103} of the open quantum dynamics that, the accuracy of \Eq{perturbation} is solely determined by the system-bath coupling $\lambda$, in great contrast to its deterministic counterpart, time-convolutionless second-order quantum master equations (TCL2) whose reliable parameter regime is yet constrained to the fast bath and the low temperature conditions\cite{Ishizaki-J.Chem.Phys.2009-p234110,Nalbach-J.Chem.Phys.2010-p194111,
Ke-J.Chem.Phys.2017-p184103,Fetherolf-J.Chem.Phys.2017-p244109}, i.e., its accuracy $\propto \lambda/(\beta\gamma^2)$. It will be verified in the coming numerical simulations whether this good quality is preserved when applied to 2D spectra.

\begin{figure*}
  \centering
  \includegraphics[width=0.9\textwidth]{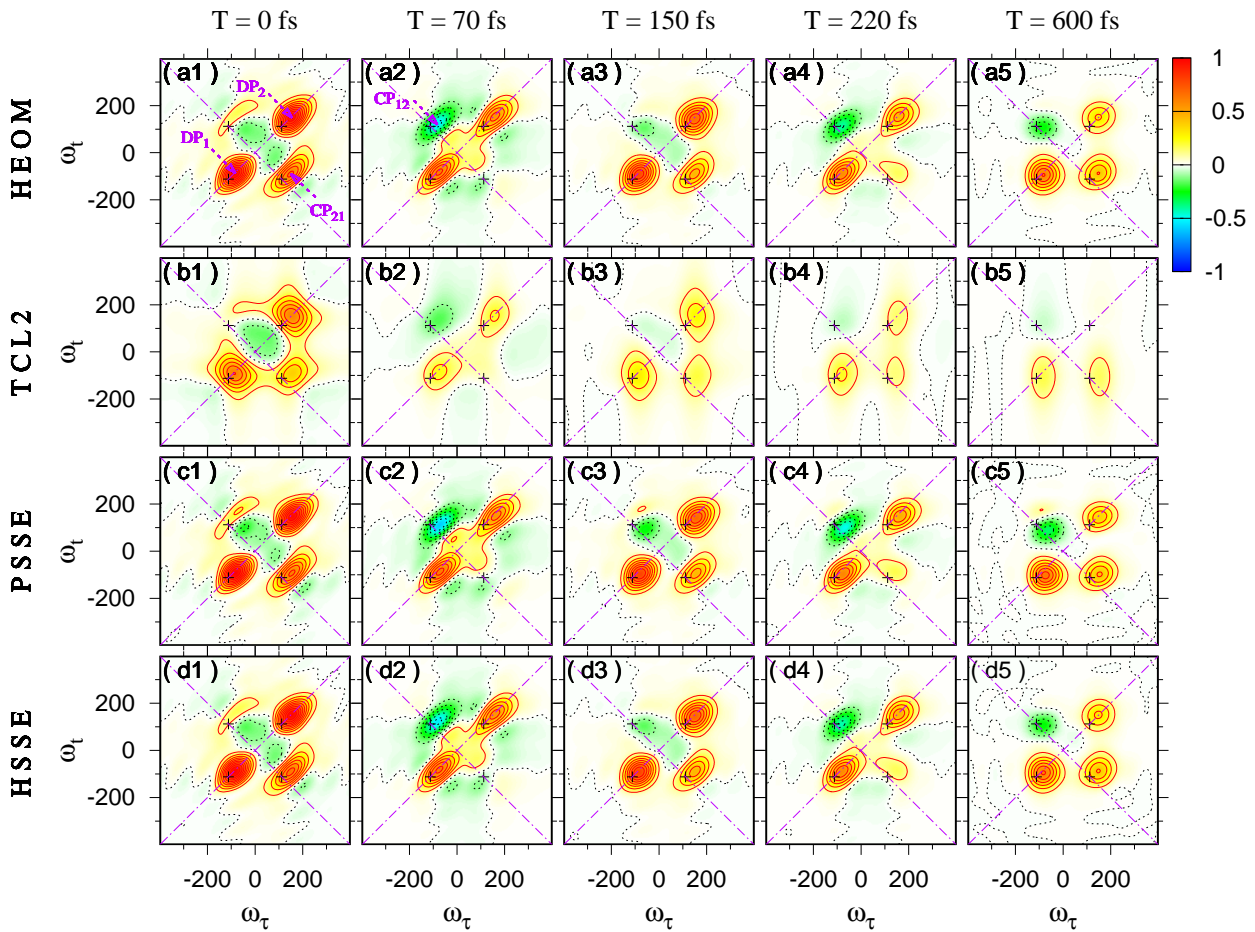}\\
  \caption{The plots of the absorptive real part of the 2D electronic spectra for an electronically coupled dimer system at the cryogenic temperature $T_{\beta}=77\,\mathrm{K}$ for various population times $T=0\,\mathrm{fs}$, $70\,\mathrm{fs}$, $150\,\mathrm{fs}$, $220\,\mathrm{fs}$, $600\,\mathrm{fs}$, with each arranged sequently from left to right. Every column from top to bottom, exhibits the results obtained through the HEOM, TCL2, PSSE, and HSSE methods. The abscissa and the ordinate are assigned by the absorption frequency $\omega_{\tau}$ and the emission frequency $\omega_t$ in units of $\mathrm{cm}^{-1}$, which are the Fourier conjugate variables  to the coherence time delay $\tau$ and rephasing time delay $t$, respectively. The positive amplitudes are drawn in yellow and red with the solid contour lines plotted in a linear scaling manner with the increment 0.1. Likewise, the negative features are in green and blue with the dashed contour lines. All the panels are scaled by the same factor which is determined by the highest peak value in 2D spectra of HEOM results at $T=0$. For the purpose of a better demonstration, the positions of three primary positive peaks $\mathrm{DP}_1$, $\mathrm{DP}_2$, and $\mathrm{CP}_{21}$ are clearly marked by magenta arrows in panel (a1), and that of the strong negative valley $\mathrm{CP}_{12}$ in panel (a2). The black criss-cross symbols in all panels display the positions related to excitonic energies $E_-=-111.8\,\mathrm{cm}^{-1}$ and $E_+=111.8\,\mathrm{cm}^{-1}$. The static disorders are not included in this case. }\label{Fig2}
\end{figure*}
\begin{figure*}
  \centering
  \includegraphics[width=0.9\textwidth]{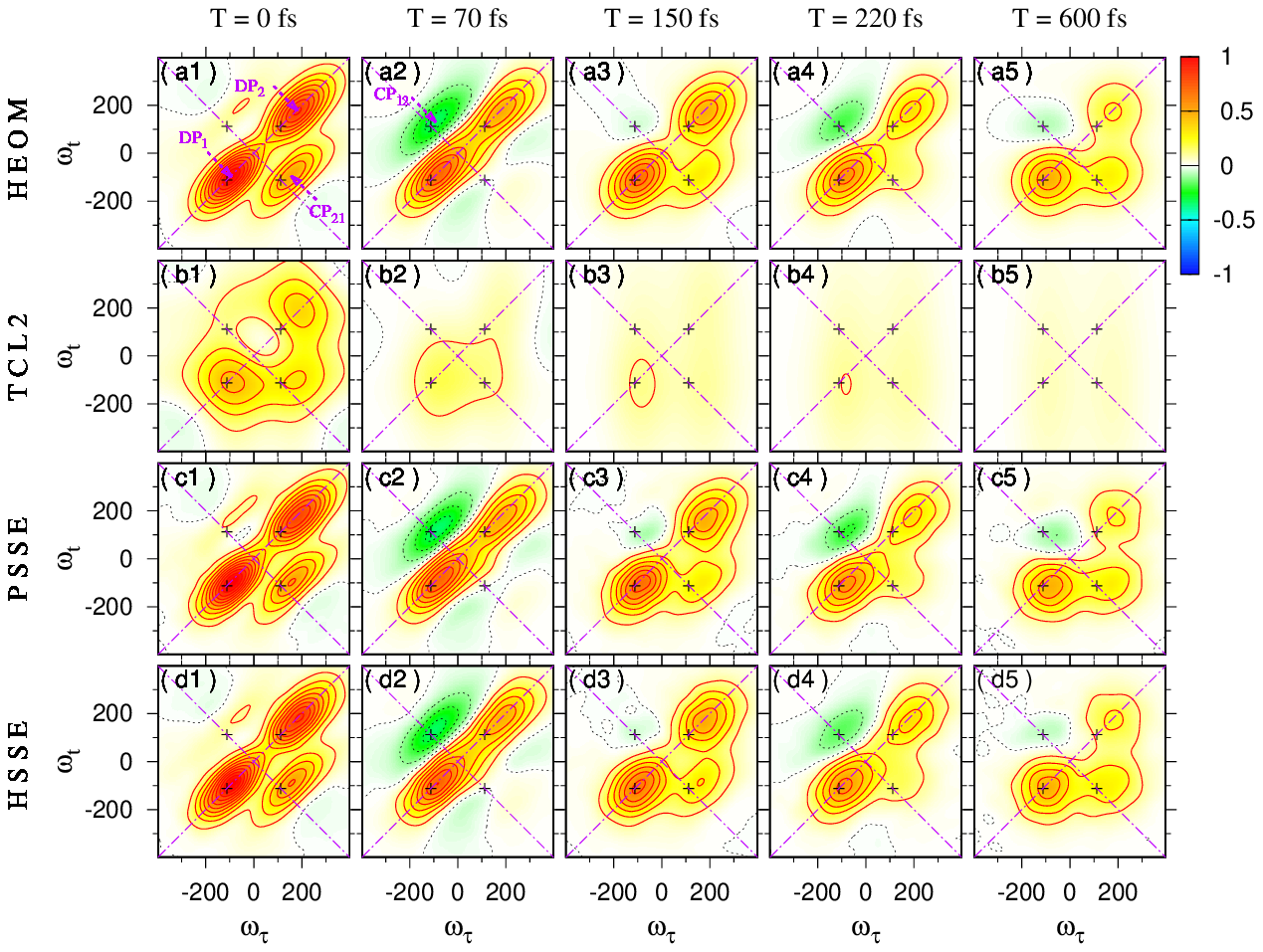}\\
  \caption{The plots of the absorptive real part of the 2D electronic spectra for the same dimer system as that in \Fig{Fig2}, except that all the simulations are carried out at the ambient temperature $T_{\beta}=300\,\mathrm{K}$. The higher temperature leads to stronger dephasing effects such that the peaks become broader to partially coalesce. The static disorders are not included in this case.
  }\label{Fig3}
\end{figure*}

To make a brief review of the strategy, one can obtain an ensemble of the forward and backward stochastic wavefunctions, i.e., $|\Psi^{L}(t,T,\tau;\bm{\xi}^L)\ra$ and $|\Psi^{R}(t,T,\tau;\bm{\xi}^R)\ra$, either being numerically exact by solving \Eq{hierarchy}, or being approximate by \Eq{perturbation} in a more efficient manner, and take their stochastic averages via \Eq{sto_response} to yield the third-order response function $\mathcal{R}(t,T,\tau)$ and afterward 2D spectra via \Eq{2D_spectra}.
For an isotropic sample consisting of an ensemble of molecules, the configurations of these molecules, their relative distances and directions, and their local environments all vary, and thereby the site energies and excitonic couplings are randomly distributed. In order to account for this effect, the static disorders are generally incorporated into the theoretical simulations by repeating the time evolution with different initial site energies and excitonic couplings that are sampled from a Gaussian distribution with the standard deviation $\sigma_s$.  Our strategy is essentially trajectory-based, such that the inclusion of static disorder is quite trivial and not much additional numerical efforts are required unless the static disorders are extremely strong. In fact, the diagonal broadening of many experimental 2D electronic spectra are primarily attributed to the static disorders, and the correlation types of these disorders can also be reflected in 2D spectra\cite{Pisliakov-J.Chem.Phys.2006-p234505}. For simplicity, in the following numerical section, we only take into account the disorders of site energies.

\section{\label{sec3}Numerical simulations}
\subsection{Test using homogeneous parallel heterodimer}
\begin{figure}
  \centering
  \includegraphics[width=0.5\textwidth]{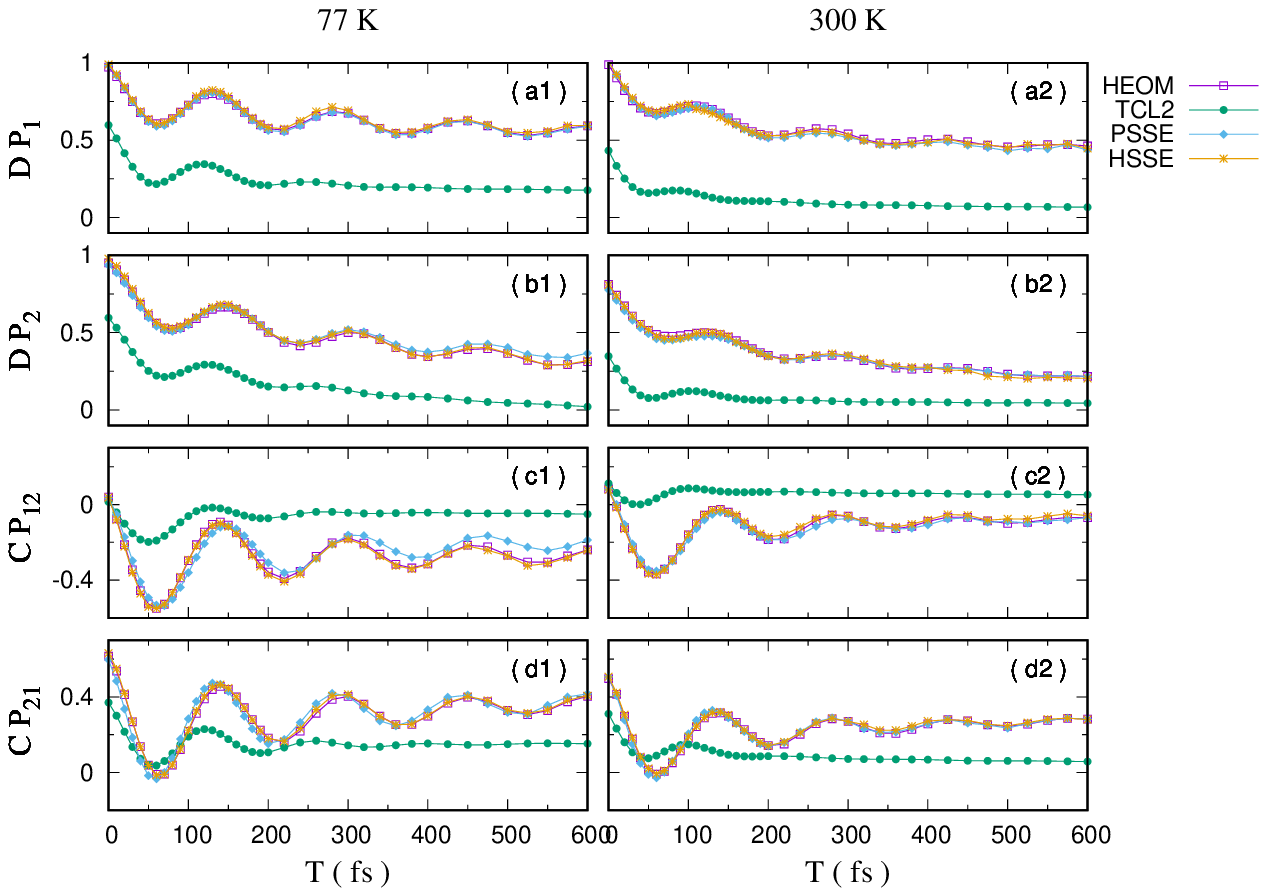}\\
  \caption{Plots of the amplitude evolution of four primary peaks ($\mathrm{DP}_1$, $\mathrm{DP}_2$, $\mathrm{CP}_{21}$, and $\mathrm{CP}_{12}$) as the function of population time delay $T$. The left column corresponds to the low temperature $T_{\beta}=77\,\mathrm{K}$, while the right to the higher temperature $T_{\beta}=300\,\mathrm{K}$. For a better comparison, the dot-line curves for a specific peak obtained through the HEOM (magenta open squares), TCL2 (green filled circles), PSSE (blue filled diamonds), and HSSE (yellow asterisks) methods are shown within the same frame. The beating periods roughly of 150fs are found in resonance with the energy splitting between two excitonic states, i.e., $\Delta_{\mathrm{gap}}=E_{+}-E_{-}=223.6\,\mathrm{cm}^{-1}$. }\label{Fig4}
\end{figure}
\begin{figure}
  \centering
  \includegraphics[width=0.5\textwidth]{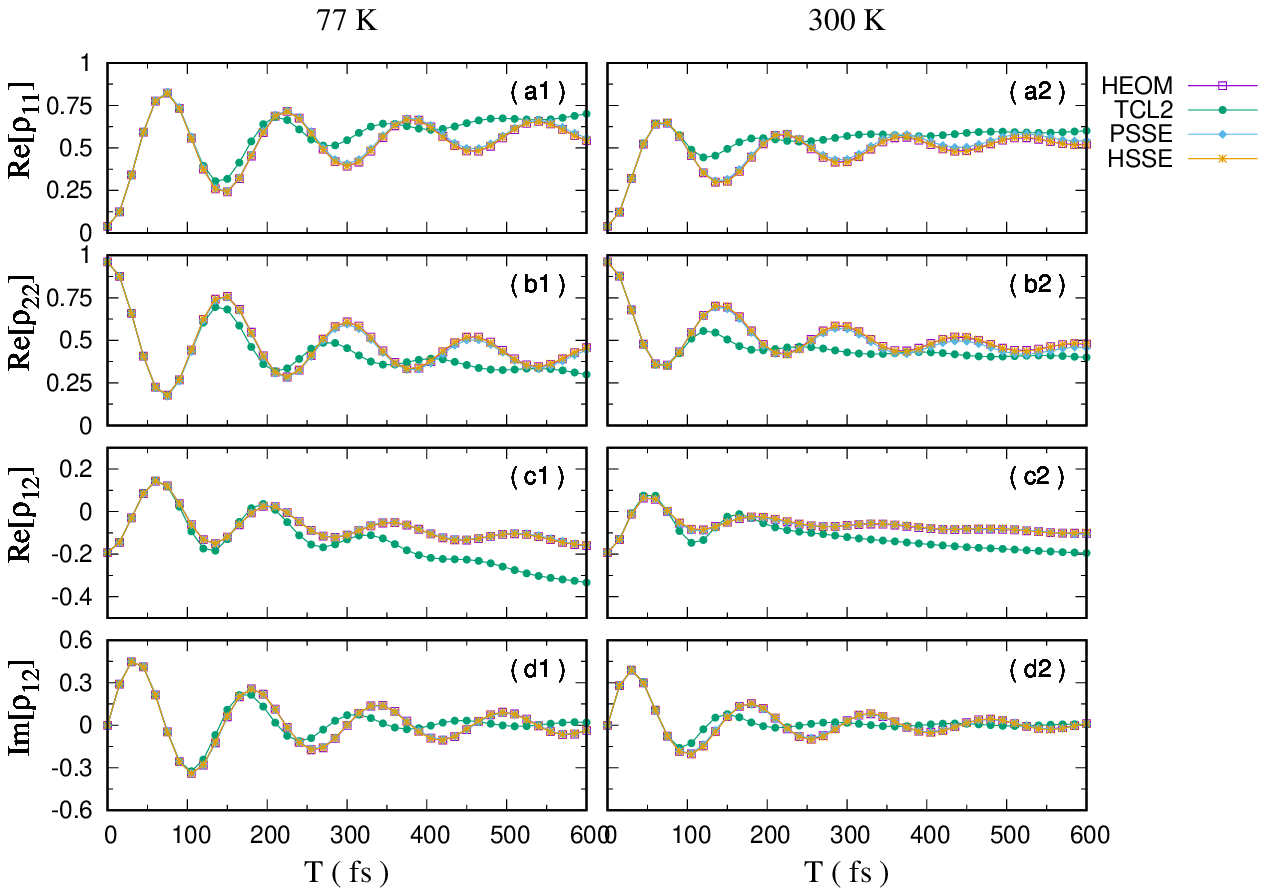}\\
  \caption{The time-evolution of the diagonal density matrix elements of the model dimer system (populations in the site representation) are shown in the first two rows, and the real and imaginary parts of the off-diagonal density matrix element $\rho_{12}$ in the last two rows, respectively. The short laser pulse excites all the excitonic states such that we prepare the initial states as $|\psi_1\ra=|\psi_2\ra=\mu_1|1\ra+\mu_2|2\ra$ to mimic the optical excitations.  }\label{Fig5}
\end{figure}
For the purpose of demonstrating the feasibility and validity of proposed schemes in calculating the 2D spectra, a dimer model system is adopted here as the test bed. Despite its simplicity, it is representative of a broad class of realistic molecular compounds\cite{Mancal-J.Chem.Phys.2004-p10556,Cho-J.Chem.Phys.2005-p114506,Halpin-Nat.Chem.2014-p196,Wang-Nat.Chem.2017-p219}, and shares some key features with more complicated complexes\cite{Zigmantas-Proc.Natl.Acad.Sci.USA2006-p12672,
Novoderezhkin-Phys.Chem.Chem.Phys.2017-p5195,Segatta-J.Am.Chem.Soc.2017-p7558}. The excitonic Hamiltonian is set as: pure electronic site energies $E_1=-50\,\mathrm{cm^{-1}}$, $E_2=50\,\mathrm{cm^{-1}}$, and the excitonic coupling $V=100\,\mathrm{cm^{-1}}$. The magnitudes of the dipole moment of two constituent monomers are $\mu_1=-0.2$, $\mu_2=1$, and they are parallelly placed such that currently it is not necessary to consider the orientational effects. The diagonalization among single excitation manifold leads to the exciton basis set composed of the lower and higher eigenstates $|-\ra=-\sin\phi|1\ra+\cos\phi|2\ra$ and $|+\ra=\cos\phi|1\ra+\sin\phi|2\ra$, with the mixing angle defined as $\phi=\frac{1}{2}\arctan{2V/(E_1-E_2)}$, which also determines the transition dipole strength redistribution. For convenience, we assume that every pigment is independently coupled to its own bath, which is characterized by Debye spectral density function shown in \Eq{Debye}, and the parameters are chosen to be $\lambda=35\,\mathrm{cm^{-1}}$ and $\gamma^{-1}=300\,\mathrm{fs}$, which are widely used in photosynthetic systems. In this case, the vertical excitation energies in \Eq{ex_hamiltonian} are specifically given as $\epsilon_1=E_1+\lambda$ and $\epsilon_2=E_2+\lambda$. As pointed out in Ref. \onlinecite{Chen-J.Chem.Phys.2010-p24505}, the inclusion of static disorders to some extent disguise the inability of approximate methods, for this reason we temporarily neglect the static disorders in this subsection by setting $\sigma_s=0$.

The absorptive real parts of 2D electronic spectra  at a sequence of  population times are plotted in \Fig{Fig2} for the cryogenic temperature $T_{\beta}=77\,\mathrm{K}$ and in \Fig{Fig3} for the ambient temperature $T_{\beta}=300\,\mathrm{K}$, obtained through four different approaches. The 2D maps obtained from the HSSE (\Eq{hierarchy}) method using 50000 trajectories are numerically converged and thus are undoubtedly in agreement with those from the HEOM method, except that the corners of \Fig{Fig2} d(5) and \Fig{Fig3} d(5) at the long population times appear to be a little ragged, which could actually be systematically alleviated by increasing random trajectory number. These results work as the benchmark, against which the reliability of perturbative methods, PSSE (\Eq{perturbation}) and TCL2 method to mine a wealth of dynamical information encoded in these 2D pictures, can be assessed. To facilitate the analysis, the positive contributions corresponding to the GSB and SE Liouville pathways, as well as the negative one from ESA pathways are given in the supporting information (SI).

There are two pronounced diagonal peaks corresponding to the lower and higher excitonic states, by convention, referred to as $\mathrm{DP}_1$ and $\mathrm{DP}_2$, respectively. Besides, a positive cross peak $\mathrm{CP}_{21}$ primarily located below the diagonal line and a negative one $\mathrm{CP}_{12}$ in the upper diagonal region, are seen conspicuously in an oscillating manner. In \Fig{Fig4}, the amplitude evolutions of these four peaks with respect to the population time delay $T$ are drawn for $T_{\beta}=77\,\mathrm{K}$ to the left and $T_{\beta}=300\,\mathrm{K}$ to the right. At $77\,\mathrm{K}$, the modulations with the period roughly of $150\,\mathrm{fs}$ ( $\approx 2\pi\Delta_{\mathrm{gap}}^{-1}$, where $\Delta_{\mathrm{gap}}=223.6\,\mathrm{cm}^{-1}$ is the energy gap between two excitonic states) are clearly observable within the whole simulation window. This oscillation is still visible at the higher temperature, though the amplitude is smaller due to stronger dephasing effects. These quantum beats in our example have a very clear physical origin. As is well known, it manifests the existence of intraband excitonic coherence state $|-\ra\la +|$ and its hermite conjugate during the population time delay $T$, which imposes a phase factor $e^{\pm i \Delta_{\mathrm{gap}}T}$ on the ESA and SE pathways. In synchronization with the amplitude oscillation, the peak shapes also exhibit periodic stretching and compression along the antidiagonal direction, as shown in \Fig{Fig2} and \Fig{Fig3}. The PSSE accurately capture these oscillating effects. Nevertheless, TCL2 predicts at most two oscillating cycles, greatly underestimating the electronic coherence lifetime. The failure of TCL2 method lies in its disability to handle with system-bath coupling in the cases where the bath responses slowly such that the non-Markovianity is extremely strong, in consistence with the previous conclusion from its application in density matrix calculations\cite{Ishizaki-Chem.Phys.2008-p185,Ke-J.Chem.Phys.2017-p184103,
Fetherolf-J.Chem.Phys.2017-p244109}. In fact, this is more straightforwardly reflected in the peak broadening mechanisms.
The environment fluctuational characteristics and their correlations are all imprinted into the peak broadening directions. For a better illustration, we supplement, in SI, the results corresponding to $\gamma^{-1}=5\,\mathrm{fs}$ approaching the Markovian limit, while other parameters are fixed. In this case, the diagonal peaks ($\mathrm{DP}_1$ and $\mathrm{DP}_2$) are elongated along the antidiagonal direction. However, as with the increase of the bath response time to $\gamma^{-1}=300\,\mathrm{fs}$, the elongation directions of these two peaks should have turned into diagonal-oriented, according to the HSSE and HEOM methods, while TCL2 results still present a qualitatively  incorrect antidiagonal broadening direction.

Owing to the introduction of stochastic noises in \Eq{stochastic} that encompass the vast majority of the system-bath coupling and the non-Markovian property, the precision of the PSSE method is far less susceptible to the bath response time, which is confirmed by the results corresponding to both $\gamma^{-1}=300\,\mathrm{fs}$ and $\gamma^{-1}=5\,\mathrm{fs}$. Provided the fact that the realistic environments are usually very large and complicated, highly probable being the mixture of both fast and slow portions. In this regard, we believe that the PSSE method is a more promising candidate than TCL2 to accurately and efficiently simulate 2D electronic spectra. Taking a global view, only a minor deviation of the PSSE results from the exact ones appears in the region near the negative $\mathrm{CP}_{12}$ peak at $77\mathrm{K}$. The emergence of this peak is mainly ascribed to the positive intersite coupling $V$, which results in the asymmetrically stronger intensity of ESA signal in the upper diagonal position, overpowering the GSB and SE signals\cite{Read-J.Phys.Chem.B2009-p6495}. On the contrary, the $\mathrm{CP}_{21}$ peak is consequently dominated by SE and GSB signals. As the time evolves ($T>0$), the energy relaxation takes place further promoting this diagonal asymmetry in the 2D spectra because the downhill transfer from higher energy state to the lower energy state is thermodynamically more favorable than the inverse process, leading to a growingly intense positive SE signal below the diagonal line ($\omega_{\tau}>\omega_t$). The better match of $\mathrm{CP}_{21}$ peak indicates that the PSSE method is fully capable to properly describe the energy transfer dynamics.  To verify this, we have also calculated the propagation of system density matrix in the site representation with the initial preparation $\mu_1^2|1\ra\la 1|+\mu_2^2|2\ra\la 2|+\mu_1\mu_2\left(|2\ra\la 1|+|1\ra \la 2|\right)$ mimicking the optical excitation by an impulsive laser pulse. It is found that not only the populations, $\rho_{11}$ and $\rho_{22}$, but also the intersite coherence $\rho_{12}$ obtained via the PSSE method are in perfect agreement with the exact results under these two parameter sets.
This, from the other perspective, reflects that the simulations of 2D electronic spectroscopy lay out an exceedingly rigorous criterium on the accuracy of theoretical methods.

Noticing that four primary peaks, $\mathrm{DP}_{1}$, $\mathrm{DP}_{2}$, $\mathrm{CP}_{12}$, and $\mathrm{CP}_{21}$, as shown in \Fig{Fig2} and \Fig{Fig3}, are obviously dislocated from the excitonic energy positions $(\omega_{\tau},\omega_t)=(E_-,E_-),\,(E_+,E_+),\,(E_-,E_+),\,(E_+,E_-)$, this behavior might stem from its slow reorganization effects from vertical excitation position to the vibrationally relaxed states,  since it is not observed in the homogeneous condition where $\gamma^{-1}=5\,\mathrm{fs}$ with the same reorganization energy. This discrepancy in 2D electronic spectra indicates that the bath characteristic frequency should play an intricate  role in the system dynamics. Therefore, in the subsection below, we will systematically investigate the influence of $\gamma$ on system dynamics and 2D peaks evolutions.
\begin{figure*}
  \centering
  \includegraphics[width=0.7\textwidth]{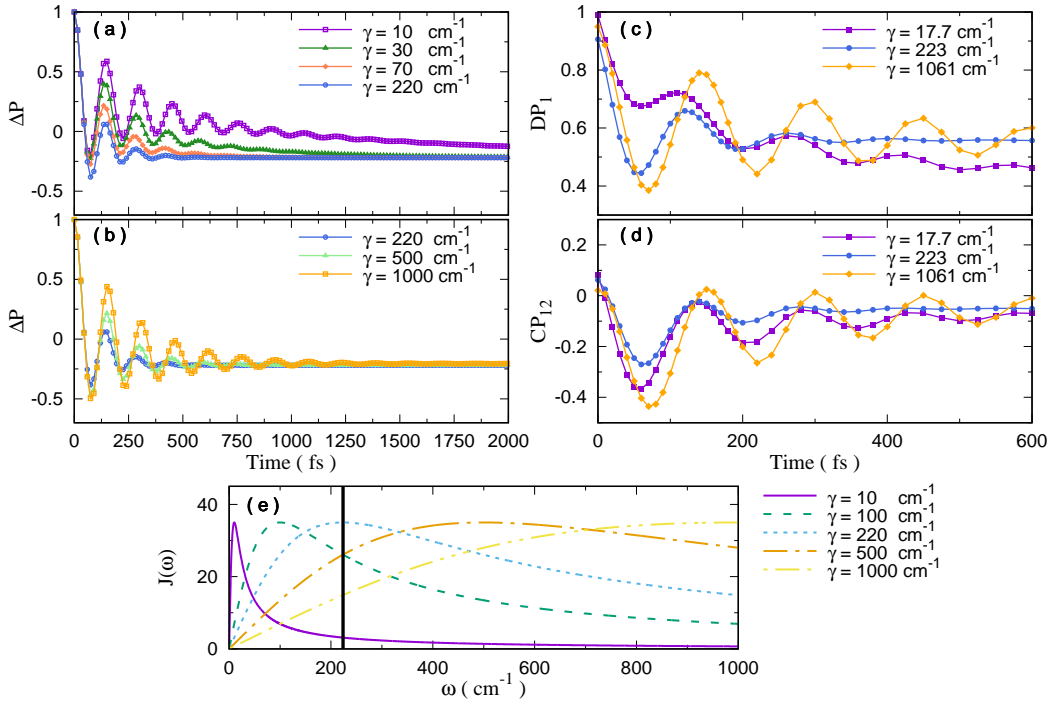}\\
  \caption{(a) and (b) plot the time-evolution of population difference $\Delta P=\rho_{22}(t)-\rho_{11}(t)$ of the dimer system with the initial condition $\rho_{22}(t=0)=1$, for various bath characteristic frequencies given in the upper right corner of each frame. (c) shows the oscillations of the diagonal peak $\mathrm{DP}_1$ in 2D spectra with respect to the population time delay T. (d) presents the oscillations of the off-diagonal peak $\mathrm{CP}_{12}$ in 2D spectra with respect to the population time delay T. $\mathrm{CP}_{12}$ is chosen as it is less prone to the population dynamics than $\mathrm{CP}_{21}$ and thus provide a more convincible probe of the coherence dynamics. All the simulations are performed at the temperature $T_{\beta}=300\,\mathrm{K}$. (e) displays the profiles of spectral density function for varying $\gamma$ and the fixed reorganization energy $\lambda=35\,\mathrm{cm}^{-1}$. }\label{Fig6}
\end{figure*}
\subsubsection{The influence of the bath characteristic frequency}
Shown in \Fig{Fig6} (a) are the plots of the population difference dynamics, $\Delta P = \rho_{22}(t)-\rho_{11}(t)$ with the initial excitation put on $\rho_{22}(t=0)=1$, for a collection of bath characteristic frequencies $\gamma$, ranging from  $10\,\mathrm{cm}^{-1}$ to $220\,\mathrm{cm}^{-1}$. Other parameters are the same as used above. When subjected to a slow bath $(\gamma=10\,\mathrm{cm}^{-1})$, the oscillation persists as long as almost $1.5\,\mathrm{ps}$. Nevertheless, if we gradually tune $\gamma$ upwards till $220\, \mathrm{cm}^{-1}$, the oscillation amplitude is correspondingly damped with a larger extent. Along with this reduction is the loss of dynamical localization effect which could be thought of as the non-Markovian features.  The stationary equilibration is achieved roughly at $T=500\,\mathrm{fs}$ when $\gamma=220\,\mathrm{cm}^{-1}$, which is nearly in resonance with the excitonic energy gap $\Delta_{\mathrm{gap}}$. More interestingly, if we keep driving $\gamma$ out of the resonance region to reach another limit, $\gamma=1000\,\mathrm{cm}^{-1}$, as shown in \Fig{Fig6} (b), the oscillation amplitudes grow back. Apparently, in contrast to the monotonic dependence on temperature and the reorganization energy, the lifetime and amplitude of the oscillations exhibit an upturned parabolic functional relationship with respect to the bath characteristic frequency $\gamma$. Noting that the density matrix has a very close connection to the 2D peaks\cite{Cheng-Chem.Phys.2007-p285}, thus, it is not surprising to find that the dephasing of the beatings is the strongest at the same turning point exactly at $223 \mathrm{cm}^{-1}$, as shown in \Fig{Fig6} (c) and (d) for $\mathrm{DP}_1$ and $\mathrm{CP}_{12}$ peaks, respectively.

In order to understand this effect, the spectral density function $J(\omega)$ for various $\gamma$ values are displayed in \Fig{Fig6} (e). Noticing that when the system-bath coupling strength, i.e., the reorganization energy $\lambda$ is fixed, the growth of bath characteristic frequency $\gamma$ would not only span the spectral density function over a larger frequency region ($\mathrm{FWHM}=2\sqrt{3}\gamma$) but also shift the maximum position $(\omega_{max}=\gamma)$, as clearly illustrated in \Fig{Fig6}(e).
When the maximal region of spectral density function coincides with the excitonic gap which is precisely marked in \Fig{Fig6}(e) as the bold black vertical line at the position $\omega=223\,\mathrm{cm}^{-1}$, the bath most effectively interacts with the system and leads to the largest decoherence rate\cite{Kjellberg-J.Chem.Phys.2006-p24106}. Put differently, if a long-lived and prominent electronic oscillation is observed even when the electron-phonon coupling is not vanishingly small and the temperature is not very low as well\cite{Engel-Nature2007-p782,Collini-Science2009-p369,Panitchayangkoon-Proc.Natl.Acad.Sci.USA2010-p12766}, the bath characteristic relaxation timescale must sit in a region far off-resonant to the excitonic energy gaps. This reinforces the argument that environment does play a vital role in regulating the electronic coherence effect.

In the photosynthetic systems where the excitonic levels are closely spaced, the overdamped environments formed by the large protein scaffolds and the surrounding matrix tend to react in a rather sluggish way and may be functionally important to the energy dissipation. On the other hand, there are also some intramolecular vibrational modes generally the high-energy end. In some studies, it is found that the distinct underdamped phonon modes being in resonance with the excitonic splittings are useful in facilitating long-range energy transfer\cite{Kolli-J.Chem.Phys.2012-p174109,Tiwari-Proc.Natl.Acad.Sci.US2012-p1203,Chin-Nat.Phys.2013-p113}. The inclusion of these discrete modes in the PSSE method is trivial, while it is also accessible in the HSSE method, as previously demonstrated in Ref. \onlinecite{Ke_J.Chem.Phys.2016_p24101} and \onlinecite{Ke-J.Chem.Phys.2017-p174105}. So it would be interesting and constructive to explore how these diverse factors cooperate to finally fulfill a highly efficient energy transfer or energy dissipation, and to what extent these features can be depicted in 2D electronic spectra.

\subsection{Non-parallel dimer with static disorders}
\begin{figure*}
  \centering
  \includegraphics[width=0.8\textwidth]{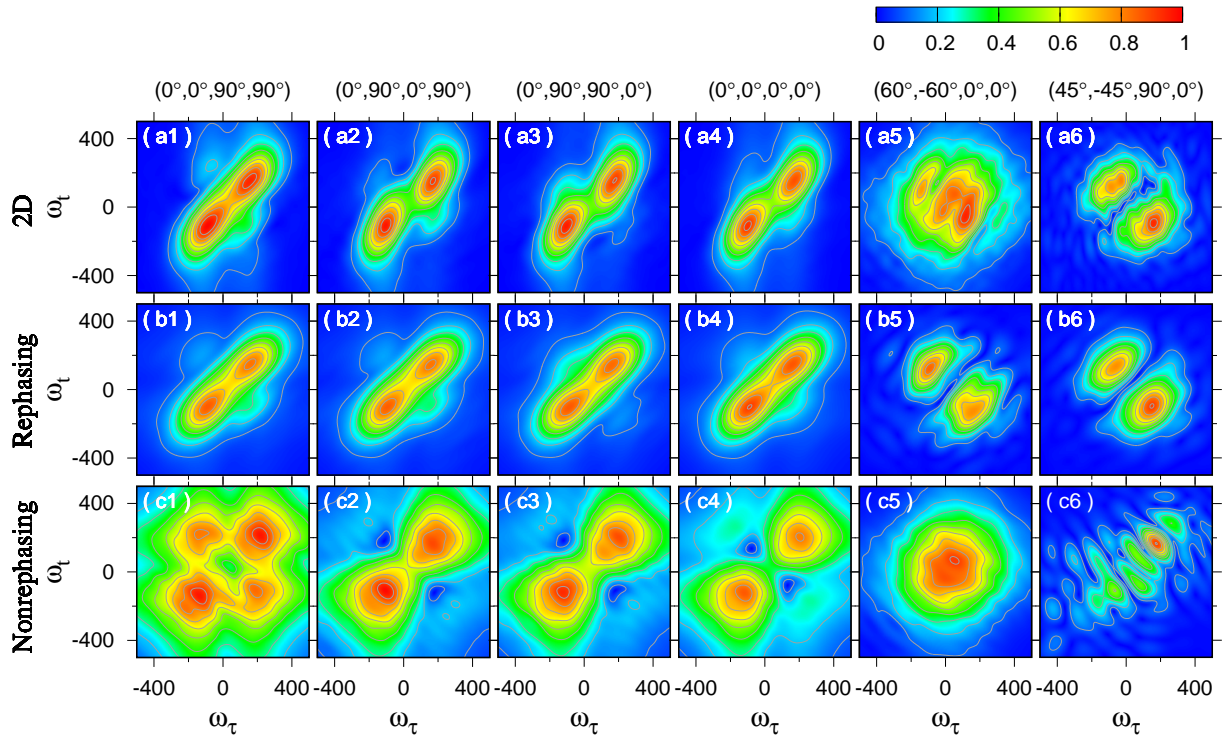}\\
  \caption{The absolute-valued 2D correlation spectra (the population time $T=0$) of an orthogonal dimer under six different pulse sequence polarization configurations. The temperature is $T_{\beta}=300K$ and the static disorders are included as well with the stand deviation  $\sigma_s=50\,\mathrm{cm}^{-1}$. The top, middle, and bottom rows are the total, rephasing part, and non-rephasing part of 2D spectra, respectively. The first three columns are three fundamental components, and their linear combinations can produce arbitrary kind of experimental polarization setup. The fourth column shows the signals corresponding to all-parallel configuration. The fifth and sixth columns are denoted as cross-peak-specific configuration and coherence-specific configuration, respectively. For the sake of better visualization, every panel is renormalized by its own maximal peak value. Since there is no guarantee about the assignment of peak signs in polarized 2D spectra, the absolute-valued 2D spectra are used. The real-valued and imaginary-valued 2D spectra are also provided in SI.}\label{Fig7}
\end{figure*}

Since the focus of the preceding subsection is mainly put on assessing the numerical performance of the proposed methods in calculating the impulsive response functions and correspondingly 2D electronic spectra, with the special attention paid to showing its superiorities when the molecules are conditioned by a slow bath, we consequently exemplify a simple dimer which is in parallel architecture and thus the rotational factor is insignificant. In this subsection, we would advance our objective toward considering a more realistic situation, that is, an non-parallel dimer with further consideration of static disorders, given that most of experiments are conducted upon a macroscopic media comprised of an ensemble of molecules experiencing various local environments and  different initial conformations. Here, the site energies of two monomers are sampled independently from a Gaussian distribution with the standard deviation $\sigma_s=50\,\mathrm{cm}^{-1}$ centering at its average energies, $E_1$ and $E_2$ given above. In particular, the dipole moments
\be
\left\{
\begin{array}{lr}
\bm{\mu}_1=&\cos\phi\,\vec{x}+\sin{\phi}\,\vec{y}\\
\bm{\mu}_2=&-\sin\phi\, \vec{x}+\cos\phi\, \vec{y}
\end{array}
\right.
\ee
are adopted in that they lead to a very special case where the transition dipole moments of the excitonic states $|+\ra$ and $|-\ra$, i.e.,
$\bm{\mu}_+$ and $\bm{\mu}_-$ are perpendicular to each other. $\vec{x}$, $\vec{y}$, and $\vec{z}$ are any three orthogonal vectors in the molecular frame, constituting a complete spatial basis set.

As mentioned before, the third-order response function is actually a fourth-rank tensor. For an isotropic sample, in view of the assumption that the angular part is separable from the other degrees of freedom,  the rotational average signal, denoted as the $\la\cdots\ra_{\mathrm{ori}}$, under the pulse polarization sequence $(\bm{e}_4\bm{e}_3\bm{e}_2\bm{e}_1)$ can be obtained by\cite{Abramavicius-Chem.Rev.2009-p2350,Schlau-Cohen-Chem.Phys.2011-p1,Yuen-Zhou-J.Chem.Phys.2011-p134505}
\be
\begin{split}
\la \mathcal{R}(t,T,\tau|\bm{e}_4\bm{e}_3\bm{e}_2\bm{e}_1)\ra_{\mathrm{ori}}
=&\sum_{\{o_4o_3o_2o_1\}} C_{o_4o_3o_2o_1}^{e_4e_3e_2e_1}\times\\
&\quad \mathcal{R}(t,T,\tau|\bm{o}_4\bm{o}_3\bm{o}_2\bm{o}_1).
\end{split}
\ee
Here $\bm{e}_i$ denote the pulse polarization directions and $\bm{e}_i\in \{\vec{X}, \vec{Y}, \vec{Z}\}$ with $\vec{X}$, $\vec{Y}$, and $\vec{Z}$ standing for three orthogonal unit vectors of the laboratory coordinate frame, and $\bm{o}_i\in \{\vec{x}, \vec{y}, \vec{z}\} $.
$\mathcal{R}(t,T,\tau|\bm{o}_4\bm{o}_3\bm{o}_2\bm{o}_1)$ means taking the results with the dipole moments in $\bm{o}_1$ direction in its interaction with the first laser pulse, and likewise for that with the second, third, and the fourth pulses.
The weighting coefficients $C_{o_4o_3o_2o_1}^{e_4e_3e_2e_1}$ are explicitly expressed as
\be
\label{rot}
\begin{split}
C_{o_4o_3o_2o_1}^{e_4e_3e_2e_1}=&\frac{1}{30}
\left[
\begin{array}{lcr}
\delta_{e_1e_2}\delta_{e_3e_4} &
\delta_{e_1e_3}\delta_{e_2e_4} &
\delta_{e_1e_4}\delta_{e_2e_3}
\end{array}\right] \times \\
&\quad \left[
\begin{array}{rrr}
4 & -1 & -1 \\
-1 & 4 & -1 \\
-1 & -1 & 4
\end{array}
\right]
\left[
\begin{array}{l}
\delta_{o_1o_2}\delta_{o_3o_4} \\
\delta_{o_1o_3}\delta_{o_2o_4}\\
\delta_{o_1o_4}\delta_{o_2o_3}
\end{array}\right].
\end{split}
\ee
In principle, there are in total $3^4=81$ kinds of the combination $\{\bm{o}_4\bm{o}_3\bm{o}_2\bm{o}_1\}$, while at most 21 terms have nonzero contributions to the final rotational average signals, and their weighting factors are specifically given by $C_{o_4o_3o_2o_1}^{e_4e_3e_2e_1}$, as shown in \Eq{rot}.

In a word, 2D spectral pattern not only depends on the molecular inherent properties but also relies heavily on the polarization setup of four laser beams. From a different perspective, it offers a powerful mean, because one can fully exploit the configurations of four laser pulses to manipulate the appearances and disappearances of the signals in a congested 2D spectra, corresponding to some specific type of Liouville pathways\cite{Zanni-Proc.Natl.Acad.Sci.USA2001-p11265,Abramavicius-Biophys.J.2008-p3613,Abramavicius-Chem.Rev.2009-p2350,
Read-Proc.Natl.Acad.Sci.USA2007-p14203,Read-Biophys.J.2008-p847,Read-J.Phys.Chem.B2009-p6495,Schlau-Cohen-Proc.Natl.Acad.Sci.USA2010-p13276,
Schlau-Cohen-Nat.Chem.2012-p389,Schlau-Cohen-IEEEJ.Sel.TopicsQuantumElectron.2012-p283}. There are, in essence, only three independent choices of the pulse polarization configurations, being $( \vec{Z}\vec{Z}\vec{X}\vec{X})$, $(\vec{Z}\vec{X}\vec{Z}\vec{X})$, and $(\vec{Z}\vec{X}\vec{X}\vec{Z})$. The absolute-valued 2D spectra corresponding to these three components are shown in the first three columns of \Fig{Fig7}, where the panels from top to bottom are the total, rephasing part and non-repahsing part of 2D signals, respectively. Any other kind of pulse sequence configuration can be obtained through the linear combination upon these three components. For instance, the signal corresponding to the most common all-parallel configuration is assembled as
\be
\begin{split}
\la \mathcal{R}(t,&T,\tau|\vec{Z}\vec{Z}\vec{Z}\vec{Z})\ra_{\mathrm{ori}}=\la \mathcal{R}(t,T,\tau|\vec{Z}\vec{Z}\vec{X}\vec{X})\ra_{\mathrm{ori}}+\\
&\la\mathcal{R}(t,T,\tau|\vec{Z}\vec{X}\vec{Z}\vec{X})\ra_{\mathrm{ori}}+
\la\mathcal{R}(t,T,\tau|\vec{Z}\vec{X}\vec{X}\vec{Z})\ra_{\mathrm{ori}},
\end{split}
\ee
which is shown in the fourth column of \Fig{Fig7}. This configuration highlights essentially the diagonal features, as can be seen clearly  from \Fig{Fig7} (a4), (b4), and (c4), whereas the cross peaks are very weak and appear to be the wings of diagonal peaks, and would probably be totally obscured if the excitonic spacings are quite small.

Inspired by the polarization shaping scheme in 2D infrared spectroscopy\cite{Zanni-Proc.Natl.Acad.Sci.USA2001-p11265}, the polarized 2D spectra in the optical regime have recently been realized owing to the prominent development of the phase stability controlling techniques, which is capable to selectively visualize some peaks of particular interest\cite{Read-Biophys.J.2008-p847,Ginsberg-Acc.Chem.Res.2009-p1352}. For instance, Read \etal had established the pulse sequence configuration $(60^{\circ},-60^{\circ},0^{\circ},0^{\circ})$ to pick out the cross peaks in the rephasing spectra of FMO complex extracted from \textit{P. phaeum} green sulfur bacteria\cite{Read-Proc.Natl.Acad.Sci.USA2007-p14203}. The signal corresponding to this configuration, called cross-peak-specific, can be obtained as
\be
\begin{split}
\la \mathcal{R}(t,T,\tau|60^{\circ},-60^{\circ},&0^{\circ},0^{\circ})\ra_{\mathrm{ori}}=3\la \mathcal{R}(t,T,\tau|\vec{Z}\vec{Z}\vec{X}\vec{X})\ra_{\mathrm{ori}}\\
&-\la\mathcal{R}(t,T,\tau|\vec{Z}\vec{Z}\vec{Z}\vec{Z})\ra_{\mathrm{ori}}.
\end{split}
\ee
The cross-peak-specific 2D spectra within our model are displayed in \Fig{Fig7} (a5), (b5), and (c5). It is confirmed that in the rephasing spectra the diagonal signals are strikingly suppressed and thus enable the clear resolution of cross-peaks. Nevertheless, all the peaks are blurred making a round and large peak in the non-rephasing spectra as this pulse polarization configuration actually cannot rule out the Liouville pathways contributing to the non-rephasing diagonal signals, and unfortunately it constitutes the major part of 2D total spectra.

Recently, Schlau-Cohen \etal have successfully devised the experiments using the pulse polarization configuration $(45^{\circ},-45^{\circ},90^{\circ},0^{\circ})$
to extract valuable structural information about the pigment-protein complex LHC\RNum{2} from green plants\cite{Schlau-Cohen-Chem.Phys.2011-p1,Schlau-Cohen-Nat.Chem.2012-p389}. The signal can be obtained theoretically as
\be
\begin{split}
\la \mathcal{R}(t,T,\tau|45^{\circ},-45^{\circ},&90^{\circ},0^{\circ})\ra_{\mathrm{ori}}=\la \mathcal{R}(t,T,\tau|\vec{Z}\vec{X}\vec{Z}\vec{X})\ra_{\mathrm{ori}}\\
&-\la\mathcal{R}(t,T,\tau|\vec{Z}\vec{X}\vec{X}\vec{Z})\ra_{\mathrm{ori}}
\end{split}
\ee
This configuration is very special. As is known, the first two pulses interact with the system to generate either the populations ($|0\ra\la 0|$, $|+\ra\la +|$, $|-\ra\la -|$), or excitonic coherence states $|\pm\ra\la \mp|$.  The second pulse is rotated to be orthogonal to the first one, such that, only the coherence states are excited. Similarly, the polarizations of the latter two pulses are also orthogonal to one another, so only the signals emitted from the coherence states are detected. As a consequence, only the Liouville pathways that the system resides in the coherence states during the population time delay contribute to the 2D electronic spectra. As such, it is possible to utilize this scheme to detect the coherence-coherence transfer which is totally a quantum effect without classical correspondence\cite{Hyeon-Deuk-J.Chem.Phys.2007-p8}. Note that the electronic coherence pathways appear in the off-diagonal position of the rephasing signals, and conversely in the diagonal position of the non-rephasing signals\cite{Cheng-J.Phys.Chem.A2008-p4254}. This explains the observations in \Fig{Fig7} (b6) and (c6). The strong non-oscillating background is removed in \Fig{Fig7} (b6) leaving two prominent and compact elliptic off-diagonal peaks. Such that, coherence-specific polarized rephasing spectra are very suitable to analyze the electronic coherence signatures.  Only the peaks along the diagonal line are discernible in the nonrephasing picture, as shown in \Fig{Fig7} (c6), but it seems to be overfeatured. Therefore, we suggest that one should be really careful about the assignment of the peaks in the coherence-specific nonrephasing spectra in the realistic experiments.  Note that in this case, 2D total signal are dominated by the rephasing part, so the cross-peaks are still well-resolved.

All these different pulse polarization configurations unveil complementarily important features about the systems, and their combinations undoubtedly constitute a flexible and powerful tool to fully interpret the elusive 2D spectra of a large host of realistic complexes.

\section{\label{sec4}Conclusion}
In summary, the strategy using the stochastic forward-backward wavefunctions, obtained by means of solving the hierarchical form or perturbative form of stochastic Schr\"odinger equations has been presented in this paper to calculate the impulsive response functions and corresponding 2D electronic spectra. The HSSE method produces numerically exact results, while the PSSE method is a good leverage between the accuracy and efficiency. Our numerical simulations based on the small dimer model system confirm, as expected, that the results acquired through the HSSE method are in complete consistence with the standard non-perturbative and non-markovian HEOM method. In addition, the perturbative prescription which is apt for arbitrary spectral density functions and capable of saving substantial computational expense in calculating 2D electronic spectra, outperforms its deterministic counterpart, the second-order time-convolutionless quantum master equations, especially when the bath response time is quite long and the temperature is not pretty low.  The newly-proposed methods, due to the following reasons, will undoubtedly serve as a powerful tool for theoretical 2D electronic spectrum simulations of realistic large-scale system, like photosynthetic complexes and organic aggregates: (1) the stochastic nature renders trivial the inclusion of static disorders which are ubiquitous in pigment-protein complexes, simply by sampling the excitonic parameters from a predetermined distribution at the beginning of every random trajectory; (2) it is extremely suitable for the highly-efficient parallel implementations; (3) the wavefunction formalism is also a great favor over the approaches based on density matrix framework, especially when applied to large-scale systems; (4) our strategy can reduce the computational overload brought by the vast doubly-excited states within the excited state absorption Liouville pathways to the least; (5) the algorithm is readily extendable to calculate higher-order optical response functions.

In the near feature, in combination with the excitonic Hamiltonians and bath parameters obtained through increasingly resolved structure and high-quality quantum chemical calculations, the proposed scheme should be very useful for making a holistic quantitative analysis of the features observed in  2D electronic spectra of photosynthetic apparatus and then clarify the key factors to the remarkably efficient energy transfer and conversion.

\section*{Supplementary Material}
See supplementary material for (1) the individual contributions of the GSB, SE, and ESA pathways to the total 2D spectra shown in \Fig{Fig1} and \Fig{Fig2}; (2) the rephasing and nonrephasing parts of the total 2D spectra shown in \Fig{Fig1} and \Fig{Fig2}; (3) the comparison of 2D spectra calculated by the HEOM, TCL2, PSSE, and HSSE methods, respectively, in the cases approaching the Markovian limits ($\gamma^{-1}$ = 5 $\mathrm{fs}$), while other parameters are the same as those used in \Fig{Fig1} and \Fig{Fig2}; (4) the supplementary real-valued and imaginary-valued 2D spectra to \Fig{Fig7}.

\section*{Acknowledgements}

The authors would like to thank the financial support from the National Science Foundation of China (Grant Nos. 21573175 and 21773191).

%

\end{document}